\documentclass[aps,pra,10pt,twocolumn,superscriptaddress,showkeys]{revtex4-1}

\usepackage{graphicx}
\usepackage{dcolumn}
\usepackage{bm}
\usepackage{amsmath}
\usepackage{amssymb}
\usepackage{amsbsy}
\usepackage{color} 
\usepackage[normalem]{ulem} 

\newcommand{\ket}[1]{\left|#1\right\rangle}

\newcommand{\ketbra}[2]{\left|#1\right>\!\!\left<#2\right|}
\newcommand{\del}{\partial}

\begin{document}

\title{Maximal Adaptive-Decision Speedups in Quantum-State Readout}

\author{B.~D'Anjou}
\author{L.~Kuret}
\author{L.~Childress}
\affiliation{Department of Physics, McGill University, Montreal, Quebec, H3A 2T8, Canada}
\author{W.A.~Coish}
\affiliation{Department of Physics, McGill University, Montreal, Quebec, H3A 2T8, Canada}
\affiliation{Quantum Information Science Program, Canadian Institute for Advanced Research, Toronto, Ontario, M5G 1Z8, Canada}
\date{\today}

\begin{abstract}
The average time $T$ required for high-fidelity readout of quantum states can be significantly reduced via a real-time adaptive decision rule. An adaptive decision rule stops the readout as soon as a desired level of confidence has been achieved, as opposed to setting a fixed readout time $t_f$. The performance of the adaptive decision is characterized by the ``adaptive-decision speedup,'' $t_f/T$. In this work, we reformulate this readout problem in terms of the first-passage time of a particle undergoing stochastic motion. This formalism allows us to theoretically establish the maximum achievable adaptive-decision speedups for several physical two-state readout implementations. We show that for two common readout schemes (the Gaussian latching readout and a readout relying on state-dependent decay), the speedup is bounded by $4$ and $2$, respectively, in the limit of high single-shot readout fidelity. We experimentally study the achievable speedup in a real-world scenario by applying the adaptive decision rule to a readout of the nitrogen-vacancy-center (NV-center) charge state. We find a speedup of $\approx 2$ with our experimental parameters. In addition, we propose a simple readout scheme for which the speedup can, in principle, be increased without bound as the fidelity is increased. Our results should lead to immediate improvements in nanoscale magnetometry based on spin-to-charge conversion of the NV-center spin, and provide a theoretical framework for further optimization of the bandwidth of quantum measurements. 
\end{abstract}

\keywords{Quantum Information, Quantum Physics}
\maketitle

\section{Introduction}

Efficient discrimination of quantum states is important for, e.g., fast qubit readout~\cite{hume2007,myerson2008,burrell2010,noek2013}, rapid feedback and steering~\cite{sayrin2011,blok2014,murch2015}, preparation of nonclassical states of light~\cite{brune2008,deleglise2008,sayrin2011,peaudecerf2014}, and nanoscale magnetometry~\cite{waldherr2012,shields2015}. Given sufficient information about the statistics and dynamics of a physical readout apparatus, it is possible to speed up a readout through a real-time adaptive decision rule (described below)~\cite{hume2007,myerson2008,burrell2010,sayrin2011,noek2013,blok2014,peaudecerf2014}. An adaptive decision rule can result in a significant reduction in the average time per measurement without any significant deterioration of the readout fidelity. Adaptive decisions can be used to improve any readout scheme, in principle, given the ability to continuously monitor the state on a time scale that is short compared to the time required for high-fidelity readout~\cite{elzerman2004,hume2007,myerson2008,barthel2009,jiang2009,morello2010,neumann2010,robledo2011,pla2013,noek2013,aslam2013,harty2014,jeffrey2014,shields2015}.

Adaptive-decision problems have a long history in probability theory. Their mathematical root can be traced back as far as the famous ``Gambler's ruin'' problem introduced by Pascal in the 17th century~\cite{edwards1983}. A general mathematical theory of adaptive decision rules, known as sequential analysis, was developed during the World War II~\cite{wald1947} and has since become a well-established part of statistical decision theory~\cite{poor1994}. Implementing an adaptive decision rule requires the ability to update a stochastically varying measure of confidence in the state, typically a likelihood function, in real time. A measurement outcome is then chosen when the confidence measure first achieves a desired value. Adaptive-decision problems are thus closely related to first-passage time analysis~\cite{risken1989,klebaner2005}; the stochastically varying confidence measure can be treated as the coordinate of a diffusing particle, which crosses a boundary when the desired confidence is reached. An adaptive decision allows for a reduction in the average measurement time, associated with a corresponding ``adaptive-decision speedup.'' Here, we formulate the adaptive decision for a two-state readout in terms of a first-passage time problem. We use this formalism to theoretically establish the maximum achievable speedups for several physical readout models. Moreover, we demonstrate that significant speedups can be achieved in practice by experimentally characterizing the adaptive-decision speedup for the detection of a NV-center charge state, effectively doubling the bandwidth of such a measurement.

The maximum achievable speedup is ultimately determined by the specific dynamics of a given readout apparatus. Some readouts may allow for a very large speedup, while others are fundamentally limited. Here, we obtain upper bounds for the adaptive-decision speedup of two commonly encountered readout schemes. The first model we consider is the paradigmatic Gaussian latching readout, in which the two states are distinguished by two noisy signals of constant but distinct intensity~\cite{kay1998,gambetta2007}. This model can be a good approximation of, e.g., a readout for the singlet and triplet states of two electron spins in a double quantum dot~\cite{barthel2009}, readouts of electron spin states based on repetitive measurement of nuclear spin ancillae~\cite{neumann2010,pla2013}, or a readout of superconducting qubits coupled to a microwave cavity~\cite{jeffrey2014}. The second model we consider is the readout based on state-dependent decay, in which the state is identified conditioned on a measured decay event. Readouts of, e.g., trapped ion qubits~\cite{noek2013}, NV-center spin qubits~\cite{robledo2011,blok2014}, and semiconductor spin qubits~\cite{elzerman2004,morello2010} can be approximated by this model. In the case of the Gaussian latching readout, we find that the speedup is bounded by a factor of $4$ as the fidelity is increased, while for the case of state-dependent decay, the speedup is bounded by a factor of $2$.

To experimentally study the achievable speedup, we implement an adaptive decision rule for a readout that discriminates between two charge states of a single NV center in diamond~\cite{aslam2013,shields2015}. The NV-center charge readout relies on the ability to distinguish strong from weak fluorescence upon illumination of the NV-center impurity. Remarkably, the NV-center charge readout can approach either of the readout models described above in distinct limits. However, for realistic experimental parameters, the dynamics of the NV-center charge readout will be intermediate between these two extremes. We therefore analyze and experimentally quantify the associated speedup. We update our level of confidence in the charge state using a quantum trajectory formalism that can easily be generalized to more complex readout schemes. We find an adaptive-decision speedup $\approx 2$ both for our experimental parameters and the experimental parameters of Ref.~\cite{shields2015}, in which a similar charge-state measurement has recently been performed. Since a shorter average measurement time increases the number of measurements that can be performed on a NV-center spin in a given time, this result should directly improve the sensitivity of magnetometry based on spin-to-charge conversion of the NV-center spin~\cite{shields2015}. A similar approach can, in principle, be applied to magnetometry using the standard spin readout of the NV center~\cite{degen2008,maze2008,taylor2008,balasubramanian2008}.

An adaptive decision can only speed up the aforementioned readout schemes by a factor of order unity. However, we show that the speedup can become parametrically large for other schemes. More precisely, we propose a simple readout, based on the discrimination of two distinct decay channels, for which the speedup becomes unbounded as the fidelity is increased. Such a readout could be realized in either atomic or quantum-dot systems. 

This article is organized as follows. In Sec.~\ref{sec:adaptiveDecisionRule}, we describe the general features of an adaptive decision rule that uses the likelihood ratio as a measure of confidence. In Sec.~\ref{sec:boundedSpeedup}, we discuss the maximal speedup for the Gaussian latching readout (Sec.~\ref{sec:gaussianLatchingReadout}) and for the readout based on state-dependent decay (Sec.~\ref{sec:stateDependentDecay}). We show that the speedup is bounded by a factor of $4$ for the Gaussian latching readout and by a factor of $2$ for the state-dependent decay readout. In Sec.~\ref{sec:chargeReadoutNV}, we introduce the charge readout of the NV center (Sec.~\ref{sec:chargeDynamics}). We describe our experimental setup as well as how the parameters of the charge dynamics were extracted from experimental data (Sec.~\ref{sec:experimentalSetup}). We then discuss how to apply the adaptive decision rule to the NV-center charge readout (Sec.~\ref{sec:errorRate}). We assess the performance of the adaptive decision rule both by performing Monte Carlo simulations based on experimental parameters and by applying the adaptive decision rule directly to experimental data. We demonstrate a factor $\approx 2$ speedup for the readout of the NV-center charge state. In Sec.~\ref{sec:parametricSpeedup}, we discuss a readout scheme where the speedup increases without bound as the readout fidelity is increased. We summarize in Sec.~\ref{sec:conclusion}. Technical details are given in Appendixes~\ref{app:gaussianAnalyticalResults}-\ref{app:experimentalData}.

\section{Adaptive decision rule \label{sec:adaptiveDecisionRule}}

The goal of readout is to discriminate between two particular states of a system of interest, $\ket{+}$ and $\ket{-}$. To achieve this, we typically let the system interact with a measurement device for a finite amount of time $t$, the readout time. During the time $t$, the measurement device records a time-resolved trajectory $\psi_t$ in the form of, e.g., an electrical or photonic signal. As illustrated in Fig.~\ref{fig:fig1}(a), a distinct trajectory is recorded when the state is either $\ket{+}$ (blue) or $\ket{-}$ (red), making it possible to discriminate between the two states. In the presence of noise, perfect discrimination is not possible. We then decide whether the state was most likely $\ket{+}$ or $\ket{-}$ based on the entire trajectory acquired during the readout time. For an optimal readout, this decision maximizes the readout fidelity $F$, namely, the probability that the state is identified correctly.

There is a tradeoff between readout fidelity and readout time; a larger fidelity is typically achieved at the cost of a longer readout time. To achieve a given fidelity target, the simplest approach is to choose a fixed readout time $t_f$ from measurement to measurement. The time $t_f$ is then selected to achieve the desired fidelity. An alternative strategy is to implement an adaptive decision rule. In this approach, each individual measurement is stopped as soon as a stopping condition is satisfied. The stopping condition is associated with reaching a minimum level of confidence in the state. A typical example of a stopping condition is illustrated in Fig.~\ref{fig:fig1}(b). When the fidelity is high, a large subset of the possible trajectories will achieve a high level of confidence in a short amount of time, while others will require substantially longer times. Therefore, the average time $T$ required for high-fidelity adaptive readout of the state may be significantly shorter than the fixed time $t_f$ required to achieve the same fidelity with a nonadaptive readout. This is the main idea underlying the field of sequential detection theory, which aims to discriminate between competing hypotheses with the shortest possible average sampling time~\cite{wald1947,bechhofer1960,tantaratana1982,poor1994}. The relative performance of the adaptive decision rule is characterized by the adaptive-decision speedup, $t_f/T$. In this work, we analyze the maximum achievable speedup for three different readout schemes listed in Table~\ref{tab:table1}. Each scheme will be described in detail in the following sections. Note that the adaptive decision rule discussed in this work does not require measurement feedback. However, it can be combined with measurement feedback to further reduce the average readout time~\cite{peaudecerf2014}.
\begin{table}
\begin{tabular}{|l||c|}
		\hline
 {\bf Readout scheme} & {\bf Maximal} $t_f/T$ \\
 \hline \hline
 Gaussian latching readout & 4 \\
 State-dependent decay & 2 \\
 Decay-channel discrimination & $\infty$ \\
 \hline
\end{tabular}
\caption{ Summary of the maximal speedups for the three readout schemes discussed in this work. \label{tab:table1}}
\end{table}

We now formalize the ideas introduced above. For a given readout time $t$, the noisy state-dependent trajectory can be represented by a function $\psi_t(t')$ defined in the interval $t' \in \left[0,t\right]$, as depicted in Fig.~\ref{fig:fig1}(a). Given the observed trajectory, we wish to decide whether the initial state was most likely $\ket{+}$ or $\ket{-}$ in a way that maximizes the fidelity $F$~\footnote{Other measures of state uncertainty, such as the entropy of the state probability distribution, may also be used when discriminating between more than two states~\cite{peaudecerf2014}.}. Alternatively, we wish to minimize the error rate $\epsilon = 1 - F$. To do so, we calculate the likelihood ratio $\Lambda_t = P(\psi_t|+)/P(\psi_t|-)$, or equivalently the log-likelihood ratio $\lambda_t = \ln \Lambda_t$. Here, $P(\psi_t|\pm)$ is the probability of observing the trajectory $\psi_t$ given that the state is initially $\ket{\pm}$. The log-likelihood ratio expresses a level of confidence in the state. When the prior probability for $\ket{\pm}$ is $P(\pm)$, the decision rule that minimizes the error rate is to choose the maximum \emph{a posteriori} estimate (MAP) of the state; if $\lambda_t > \lambda_{\textrm{th}}$ ($\lambda_t < \lambda_{\textrm{th}}$), we decide that the initial state was most likely $\ket{+}$ ($\ket{-}$). Here, $\lambda_{\textrm{th}} = \ln\left[P(-)/P(+)\right]$ is the optimal decision threshold. When the prior probabilities are assumed to be equal, $P(+)=P(-)$ and $\lambda_{\textrm{th}}=0$, the decision reduces to the maximum-likelihood estimate (MLE). In most of this article, we assume equal prior probabilities so that the MLE is optimal. We briefly discuss unequal prior probabilities and the MAP in Sec.~\ref{sec:errorsBias}. 

When using a nonadaptive decision rule, we fix $t=t_f$ for each measurement and base our decision on the value of $\lambda_{t_f}$ (nonadaptive MLE or MAP). When using an adaptive decision rule, in contrast, we stop each measurement as soon as $\lambda_t \ge \lambda_+$, $\lambda_t \le \lambda_-$ or $t \ge t_M$, as illustrated in Fig.~\ref{fig:fig1}(b). Here, $\lambda_{\pm}$ are stopping thresholds and $t_M$ is a maximum readout time. We then base our decision on the value of $\lambda_t$ at the stopping time (adaptive MLE or MAP). The average value of $t$ from measurement to measurement gives the average readout time $T$. For each of the three readout schemes listed in Table~\ref{tab:table1} , we have proven that a symmetric choice of stopping thresholds, $\lambda_+ = -\lambda_- = \bar{\lambda}$, minimizes $T$ for a given error rate $\epsilon$ when the states are equally likely. For the readout discussed in Sec.~\ref{sec:chargeReadoutNV}, we optimize the stopping thresholds numerically.

\section{Bounded speedup \label{sec:boundedSpeedup}}

In this section, we apply the ideas of Sec.~\ref{sec:adaptiveDecisionRule} to two commonly encountered readout schemes, the Gaussian latching readout and the readout based on state-dependent decay. Although these schemes are usually not perfectly realized in practice, they are relevant to understand the limitations of a wide variety of readouts. In particular, both models can be mapped to extreme limits of the NV-center charge readout discussed in Sec.~\ref{sec:chargeReadoutNV}.

\subsection{Gaussian latching readout \label{sec:gaussianLatchingReadout}}

\begin{figure}
\centering
\includegraphics[width=\columnwidth]{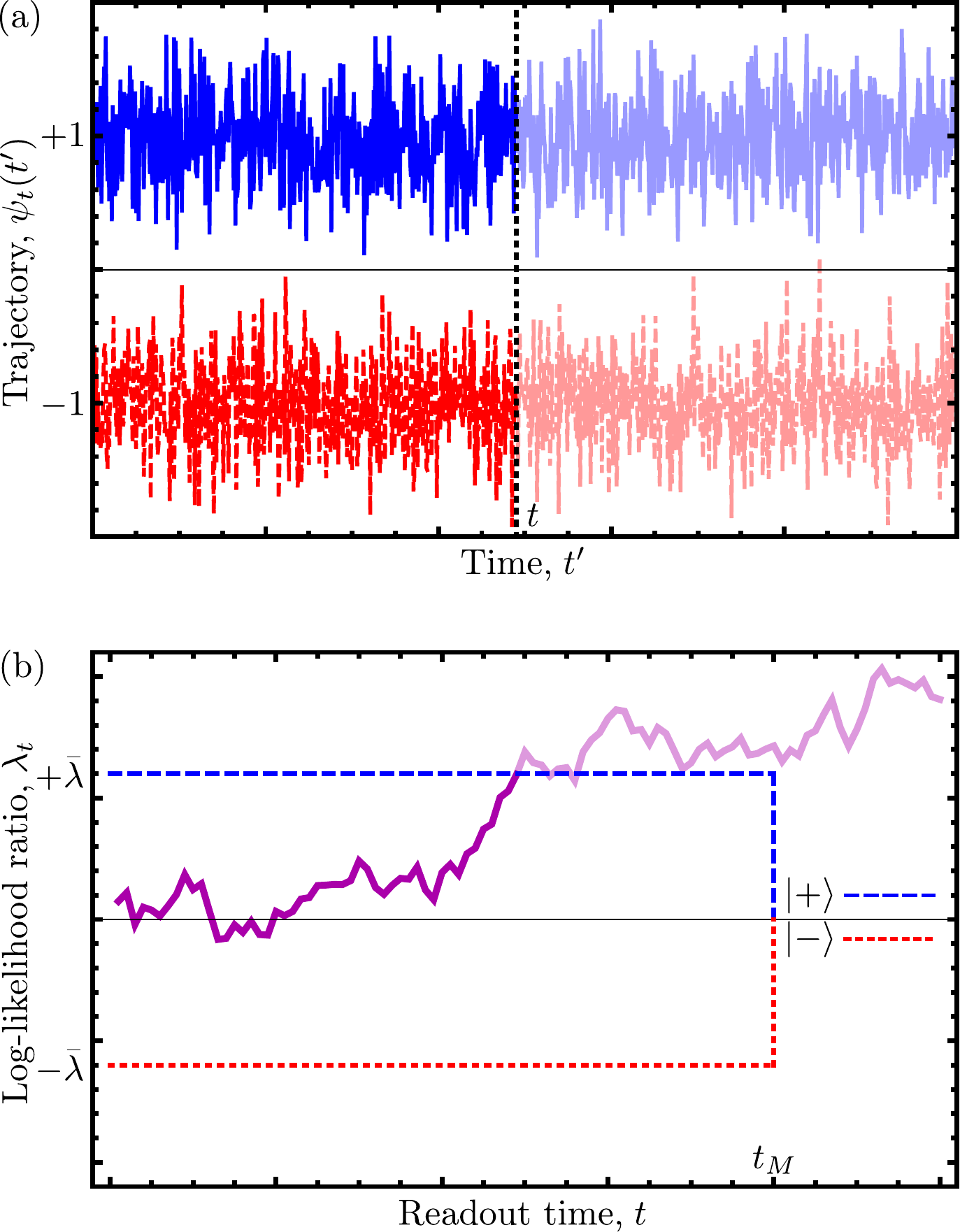}
\centering
\caption{(a) Readout trajectories for the Gaussian latching readout. The trajectory for state $\ket{+}$ (solid blue) has an average of $+1$, while the trajectory for state $\ket{-}$ (dashed red) has an average of $-1$. Both trajectories have the same signal-to-noise ratio per unit time $r$. The identification of the state is made based on the entire trajectory acquired during the readout time $t$ (dotted black). (b) Schematic representation of the adaptive decision rule for the Gaussian latching readout. As soon as the log-likelihood ratio $\lambda_t$ (solid magenta) satisfies the stopping condition $\lambda_t \ge \bar{\lambda}$, $\lambda_t \le -\bar{\lambda}$ or $t \ge t_M$, data acquisition is stopped. The sign of $\lambda_t$ is then used to choose the qubit state (dashed blue for $\ket{+}$ and dotted red for $\ket{-}$). Here, we choose $\ket{+}$. \label{fig:fig1}}
\end{figure}

The Gaussian latching readout is the simplest, most widespread and most tractable model of state discrimination~\cite{kay1998}. In this scheme, each state $\ket{\pm}$ gives rise to a state-dependent trajectory with constant average $\pm 1$ and subject to Gaussian white noise, as illustrated in Fig.~\ref{fig:fig1}(a). A latching readout is characterized by the absence of state relaxation at all times. Even though most readouts are limited by state relaxation~\cite{gambetta2007,danjou2014,khezri2015}, many implementations will approximately become latching readouts as the error rate is decreased~\cite{barthel2009,neumann2010,pla2013,jeffrey2014,shields2015}. 

Formally, the probability distributions for $\psi_{t}(t')$ conditioned on the state are
\begin{align}
	P(\psi_{t}|\pm) = A \exp{\left\{-\frac{r}{2} \int_0^t dt'\, [\psi_t(t')\mp 1]^2\right\}}. \label{eq:signalDistribution}
\end{align}
Here, $r$ is the power signal-to-noise ratio per unit time and $A$ is a normalization constant independent of the state. From Eq.~\eqref{eq:signalDistribution}, it is straightforward to show that the log-likelihood ratio is $\lambda_t = 2 r \int_0^t dt'\, \psi_t(t')$. Thus, in the case of the Gaussian latching readout, a maximum-likelihood decision can be made by integrating the trajectory up to time $t$.

If the trajectory is acquired over the fixed interval $\left[0,t_f\right]$, the error rate takes the form~\cite{gambetta2007}:
\begin{align}
	\epsilon = \frac{1}{2} \mathrm{erfc}{\left(\sqrt{\frac{r\,t_f}{2}}\right)}, \label{eq:errorNonAdaptive}
\end{align}
where $\mathrm{erfc}$ is the complementary error function~\cite{gradshteyn2007}. A derivation of Eq.~\eqref{eq:errorNonAdaptive} is included as part of Appendix~\ref{app:gaussianAnalyticalResults}. Alternatively, we can implement an adaptive decision rule. Because of the symmetry of the problem under interchange of $\ket{+}$ and $\ket{-}$, it is optimal to choose symmetric stopping thresholds. Therefore, we stop the readout as soon as $\lambda_t \ge \bar{\lambda}$, $\lambda_t \le -\bar{\lambda}$ or $t \ge t_M$. As in the nonadaptive case, we choose the state according to the sign of $\lambda_t$ at the end of the sequence. The adaptive decision rule is illustrated in Fig.~\ref{fig:fig1}(b). The error rate $\epsilon$ and average readout time $T$ are then defined parametrically as a function of both $\bar{\lambda}$ and $t_M$. The calculation of $\epsilon$ and $T$ can be recast as a first-passage time problem~\cite{risken1989,klebaner2005}, as detailed in Appendix~\ref{app:gaussianAnalyticalResults}. For clarity, here we only give the result for the case when the probability of reaching $t \ge t_M$ is negligible, $r\,t_M \gg \min{\left(1,\bar{\lambda}^2\right)}$ (see Appendix~\ref{app:gaussianAnalyticalResults}). The corresponding error rate and average readout time are given by
\begin{align}
	\epsilon = \frac{1}{1 + e^{\bar{\lambda}}}, \;\;\; T = \frac{\bar{\lambda}}{2r}\, \tanh{\left(\frac{\bar{\lambda}}{2}\right)}. \label{eq:errorAdaptive}
\end{align}
By varying the stopping threshold $\bar{\lambda}$, we can map the functional relation $\epsilon(T)$. As $\epsilon \rightarrow 0$, Eq.~\eqref{eq:errorAdaptive} implies that $\epsilon \approx \exp{(-2 r\,T)}$. In the nonadaptive case, Eq.~\eqref{eq:errorNonAdaptive}, we instead have $\epsilon \approx \exp(-r\,t_f/2)/\sqrt{2\pi r\,t_f}$. Therefore, in the limit of small $\epsilon$, the adaptive decision rule asymptotically achieves the same error rate at an average time 4 times shorter than the nonadaptive decision rule~\cite{bechhofer1960,poor1994}. Thus, the maximal speedup is given by
\begin{align}
\frac{t_f}{T} \sim 4,\;\;\; (\epsilon \rightarrow 0).
\end{align}
Here and throughout this article, ``$\sim$'' indicates an asymptotic equality.

\subsection{State-dependent decay \label{sec:stateDependentDecay}}

\begin{figure}
\centering
\includegraphics[width=\columnwidth]{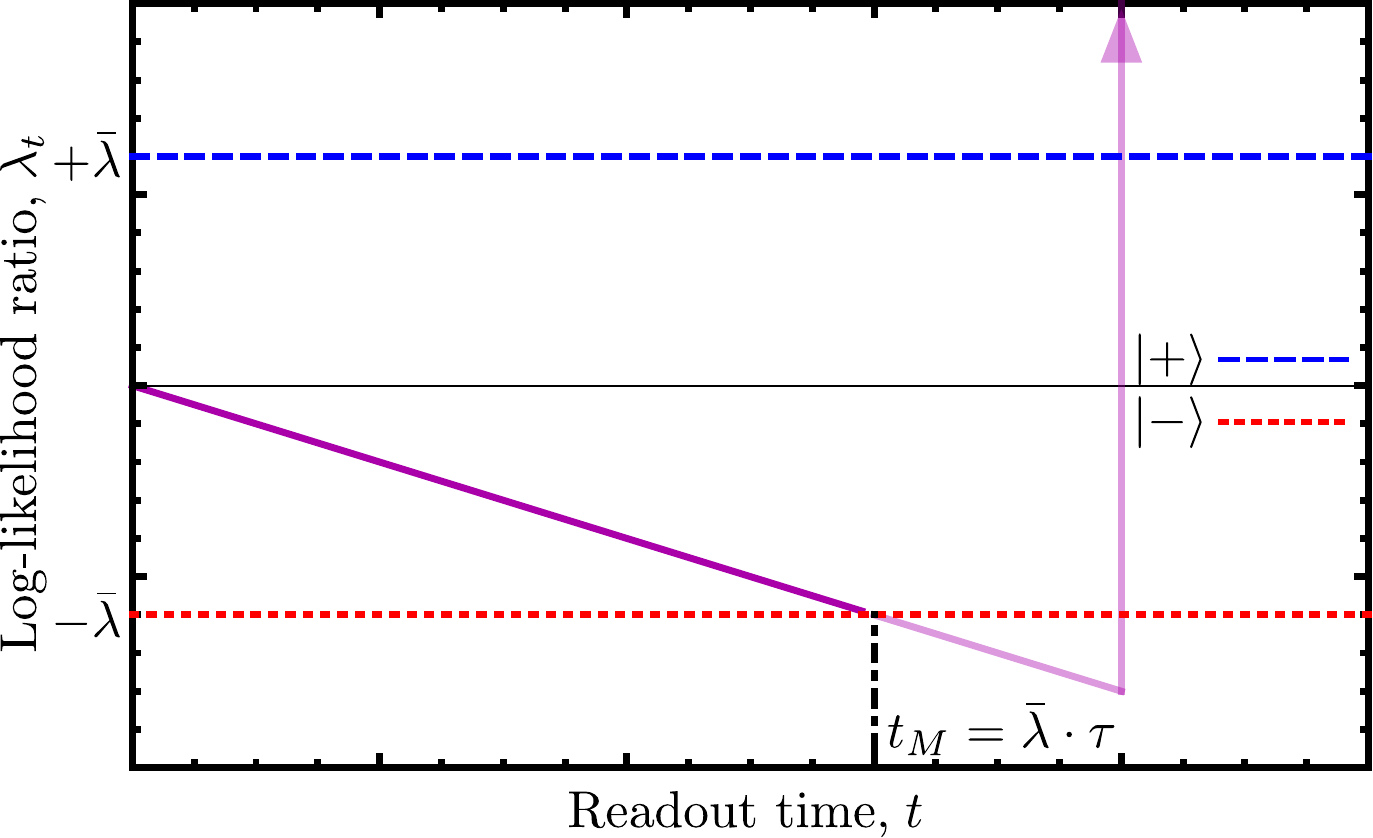}
\centering
\caption{Adaptive decision rule for the state-dependent decay readout. The readout is stopped as soon as the likelihood ratio $\lambda_t$ (solid magenta) satisfies $\lambda_t < -\bar{\lambda}$ or $\lambda_t > \bar{\lambda}$. In this case, this pair of conditions is equivalent to stopping the readout as soon as an event is detected before a maximum time $t_M = \bar{\lambda}\,\tau$ (dot-dashed black). We choose the state according to the sign of $\lambda_t$ at the stopping time (dashed blue for $\ket{+}$ and dotted red for $\ket{-}$). In this example, we choose $\ket{-}$. For the magenta trajectory shown above, this choice results in an error since an event occurs after time $t_M$, as indicated by the sudden divergence in $\lambda_t$. \label{fig:fig2}}
\end{figure}

As a second example, we consider the class of readouts relying on state-dependent decay. In these schemes, the state $\ket{+}$ decays to $\ket{-}$ with probability per unit time $\tau^{-1}$. The detection of a decay event, e.g., the detection of an emitted photon or of a tunneling electron, indicates that the initial state was $\ket{+}$. In contrast, the absence of a decay event indicates that the state was $\ket{-}$. This situation is approximately realized in, e.g., trapped ion qubits~\cite{noek2013}, NV-center spin qubits~\cite{robledo2011,blok2014}, and semiconductor spin qubits~\cite{elzerman2004,morello2010}. To facilitate the discussion, we assume that the events are detected with perfect efficiency. Moreover, we assume that a detected event is always the result of a transition; i.e., there are no ``dark'' counts. In this case, the log-likelihood ratio is $\lambda_t = - t/\tau$ when no event has been detected after a readout time $t$ because the detection probability of a decay event decreases exponentially with time. When an event is detected, $\lambda_t$ suddenly becomes infinite because an event is guaranteed to have resulted from the initial state being $\ket{+}$. This scenario is illustrated in Fig.~\ref{fig:fig2}.

When using a nonadaptive decision rule, we wait for a time $t_f \gtrsim \tau$ and base the decision on whether a decay event is detected in the interval $\left[0, t_f \right]$, as indicated by the sign of $\lambda_t$. The error rate is given by the probability that the event occurs after time $t_f$ assuming an initial state $\ket{+}$, $e^{-t_f/\tau}$, multiplied by the probability $1/2$ that the state was initially $\ket{+}$:
\begin{align}
	\epsilon = \frac{1}{2} e^{-t_f/\tau}. \label{eq:errorEmission}
\end{align}

In contrast with the nonadaptive case, the adaptive decision rule stops the readout as soon as $\lambda_t < -\bar{\lambda}$ or $\lambda_t > \bar{\lambda}$, as shown in Fig.~\ref{fig:fig2}. Note that, for this particularly simple readout scheme, the use of a maximum readout time $t_M$ is redundant since the above stopping condition is the same as stopping the readout as soon as an event is detected before a maximum time $t_M = \bar{\lambda}\,\tau$. Variations of this adaptive decision rule have been implemented in Refs.~\cite{noek2013,blok2014}. The error rate is now given by the probability of an event occurring after time $t_M$. To achieve the same error rate as the nonadaptive decision rule, Eq.~\eqref{eq:errorEmission}, we thus set $t_M = t_f$. However, we find that the readout time is now given, on average, by $(1-e^{-t_f/\tau})\tau $ when the state is $\ket{+}$. Thus, the average readout time is $T = \left[(1-e^{-t_f/\tau})\tau + t_f\right]/2$ because, as noted in Ref.~\cite{noek2013}, we must always wait for the full duration $t_f$ when the state is $\ket{-}$. Thus, once again, we find a bounded speedup in the limit of vanishing error rate, $t_f \rightarrow \infty$:
\begin{align}
	\frac{t_f}{T} \sim 2,\;\;\;(\epsilon\rightarrow 0). \label{eq:decaySpeedup}
\end{align}

\section{Experiment: NV-center charge readout \label{sec:chargeReadoutNV}}

The analysis of Sec.~\ref{sec:boundedSpeedup} provides limits on the achievable speedup for two idealized readout mechanisms. Many experimental realizations of a two-state readout can approach one or the other model in different limits, but under typical experimental conditions neither limit can be strictly realized. We must therefore use an algorithm for the adaptive decision rule suited to the underlying readout mechanism. 

Here, we provide an explicit implementation of an adaptive decision rule for experimental two-state readout in a regime that lies intermediately between the two idealized examples of Sec.~\ref{sec:boundedSpeedup}. Specifically, we examine fluorescence-based charge-state detection of the NV center in diamond. As explained below, this system approaches a Gaussian latching readout in the limit of high fluorescence rates and long charge-state relaxation times, while it can begin to resemble state-dependent decay for very low excitation powers. In the following, we describe the main features of the NV charge readout system and use experimental data to extract the underlying system parameters. We then both simulate and experimentally implement the adaptive decision rule with the help of a simple algorithm that uses our knowledge of the dynamics to update the likelihood ratio in real time. For our experimental parameters, we find a speedup of $t_f/T \approx 2$. The resulting improvement in detection bandwidth can improve the sensitivity of an NV-center magnetometer using spin-to-charge conversion~\cite{shields2015}.

\subsection{Charge dynamics of the NV center in diamond \label{sec:chargeDynamics}}

The NV center in diamond is typically observed in two charge states, namely, the negatively charged NV$^{-}$ and the neutral NV$^{0}$. In keeping with our notation, we label NV$^{-}$ and NV$^{0}$ by $\ket{+}$ and $\ket{-}$, respectively. When the NV$^{-}$ impurity is illuminated with yellow light ($\approx 594\,\textrm{nm}$), transitions between the NV$^{-}$ ground and excited states scatter photons which are detected at a rate $\gamma_+$. The frequency of the yellow light is below the minimum threshold required to resonantly excite optical transitions of NV$^{0}$, and thus only a residual detection rate $\gamma_{-} \ll \gamma_{+}$ is observed in the neutral state. Note that the rates $\gamma_{\pm}$ include the effect of dark counts and imperfect detection efficiency. This difference in fluorescence rates enables direct readout of the NV-center charge state~\cite{aslam2013,shields2015}. In addition, the incident light can cause ionization (recombination) of the NV$^{-}$ (NV$^{0}$) state with probability per unit time $\Gamma_+$ ($\Gamma_-$). As illustrated in Fig.~\ref{fig:fig3}(a), this leads to an alternating process of ionization and recombination that limits readout fidelity~\footnote{A similar process limits the readout fidelity of fluorescence-based readouts of trapped atoms and ions~\cite{gehr2010,noek2013,wolk2015}}. Typically, the condition $\min(\gamma_{+},\gamma_{-}) \gg \max(\Gamma_{+},\Gamma_{-})$ is satisfied, so many transitions occur between ionization and recombination events. The trajectory $\psi_t$ can therefore be modeled as a hidden two-state Markov process subject to state-dependent Poissonian noise of intensity $\gamma_{\pm}$~\cite{aslam2013,shields2015}. Because of detector bandwidth limitations, the maximum readout time $t_M$ is usually separated into $N$ bins of duration $\delta t = t_M/N$, so we observe $\delta n_i$ photons in bins $i = 0,1,2, \,\dots\, ,N-1$. Thus, the measured trajectory can be represented as an $N$-component vector $\psi_N = (\delta n_{N-1}, \,\dots \, ,\delta n_0)$.

Note that the charge readout of the NV center approaches the two cases discussed in Sec.~\ref{sec:boundedSpeedup} in different limits. When $\min{(\gamma_+, \gamma_-)} \gg \delta t^{-1}$, the noise statistics of the NV-center charge readout are approximately Gaussian, albeit with asymmetric signal-to-noise ratios per unit time:
\begin{align}
	r_\pm = \frac{(\gamma_+ - \gamma_-)^2}{4 \gamma_{\pm}}.
\end{align}
To perform a high-fidelity readout, the rate of information gain must be faster than state relaxation, $\min{(r_+,r_-)} \gg \max{(\Gamma_+,\Gamma_-)}$. For an adaptive advantage, the rate of information gain must also be slower than the sampling rate, $\max{(r_+ , r_- )} \ll \delta t^{-1}$. Combined with $\min{(\gamma_+,\gamma_-) \gg \delta t^{-1}}$, the last inequality implies the necessary condition $\max{(r_+,r_-)} \ll \min{(\gamma_+,\gamma_-)}$ or $|\gamma_+ - \gamma_-| \ll 2\min{(\gamma_+,\gamma_-)}$. We conclude that when state relaxation is negligible and when $\gamma_+$ and $\gamma_-$ are both large and comparable in magnitude, the NV-center charge readout is approximately described by the model of Sec.~\ref{sec:gaussianLatchingReadout} with signal-to-noise ratio $r\approx r_+ \approx r_-$. In contrast, when $\gamma_+ \gg \max(\gamma_-,\Gamma_+,\Gamma_-)$, the NV-center charge readout can be approximately modeled by the state-dependent decay of Sec.~\ref{sec:stateDependentDecay} with $\tau = \gamma_+^{-1}$. Indeed, in this limit the presence or absence of a single photon on a time scale of a few $\gamma_+^{-1}$ is sufficient to choose the state with high confidence. Therefore, the detection of additional photons adds little information, and the state-dependent decay model is a good approximation. For our parameters, the experiments described below do not fall strictly into either of these limits. It is therefore interesting to characterize the speedup in this intermediate regime.

\subsection{Experimental setup and data \label{sec:experimentalSetup}}

\begin{figure}
\centering
\includegraphics[width=\columnwidth]{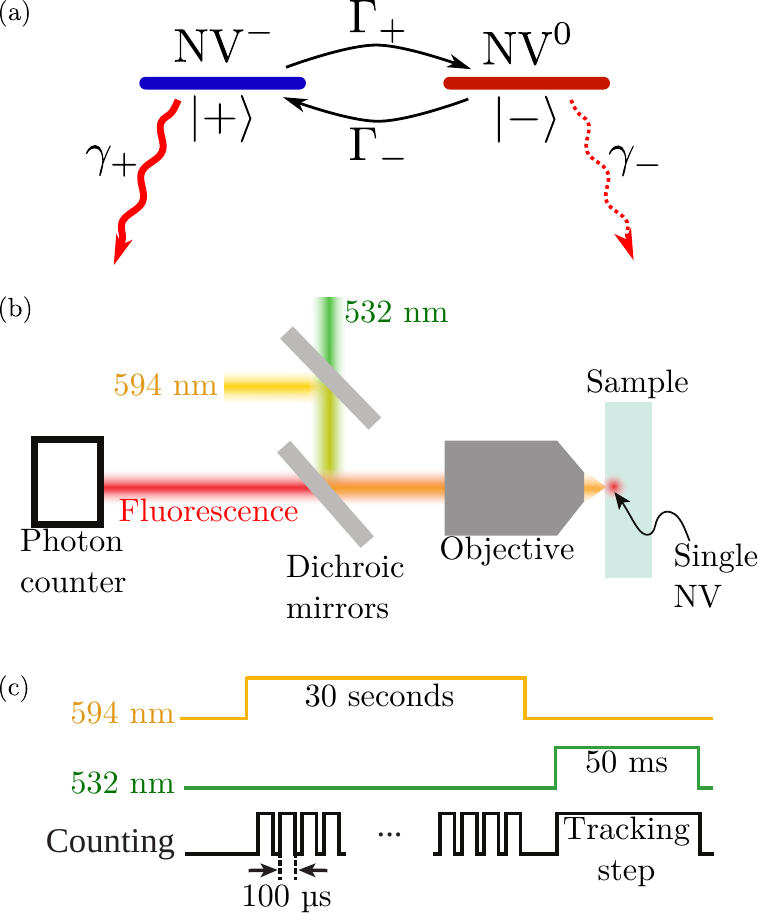}
\centering
\caption{Schematic representation of (a) the NV-center charge dynamics described in Sec.~\ref{sec:chargeDynamics}, (b) the experimental apparatus and (c) the experimental sequence. \label{fig:fig3}}
\end{figure}

We characterize the performance of the adaptive decision rule using experimental data. The data were acquired using a home-built confocal microscope with $594$-$\textrm{nm}$ excitation. The $594$-$\textrm{nm}$ wavelength lies in between the zero-phonon line of the NV$^{-}$ ($637\,\textrm{nm}$) and NV$^{0}$ ($575\,\textrm{nm}$), so it only efficiently excites the negatively charged state. The excitation light is focused through a high numerical aperture objective (NA 1.35) onto a $\left<111\right>$ cut chemical-vapor-deposition-grown diamond in which single defects can be resolved. Photons emitted in the wavelength range $645-800\;\textrm{nm}$ are collected and detected with a single-photon counter. This detection range overlaps strongly with the NV$^{-}$ fluorescence spectrum and only weakly with the NV$^{0}$ fluorescence, while rejecting Raman scattering from the diamond. The experimental setup is illustrated schematically in Fig.~\ref{fig:fig3}(b).

\begin{figure}
\centering
\includegraphics[width=\columnwidth]{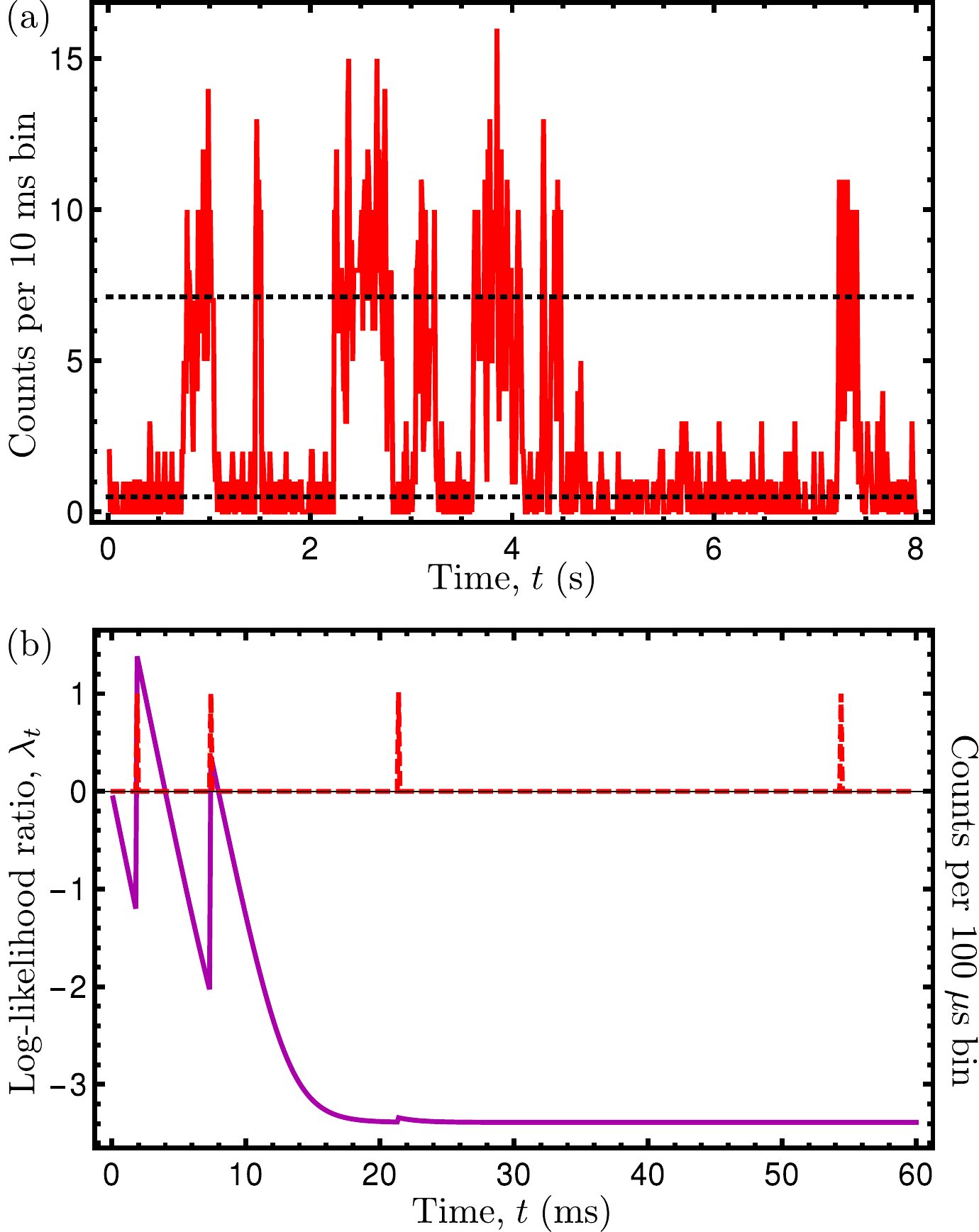}
\centering
\caption{(a) Experimental photon-count trajectory illustrating the state-dependent fluorescence rate and two-level switching of the NV-center charge state for an illumination power of $0.55\,\mu\textrm{W}$. Here, the photon detection rate fluctuates between $\gamma_+ \approx 720 \, \textrm{Hz}$ for NV$^{-}$ and $\gamma_- \, \approx 50\,\textrm{Hz}$ for NV$^{0}$, as discussed in Sec.~\ref{sec:experimentalSetup}. The corresponding average counts per $10$-$\textrm{ms}$ bin are indicated by the horizontal dotted black lines. The NV-center ionization and recombination rates are found to be $\Gamma_+ \approx 3.6 \,\textrm{Hz}$ and $\Gamma_- \approx 0.98 \,\textrm{Hz}$, respectively. (b) Log-likelihood ratio $\lambda_t$ (solid magenta) and counts per $100$-$\mu\textrm{s}$ bin (dashed red) for the first $60\,\textrm{ms}$ of the data shown in (a). The log-likelihood ratio was calculated using Eq.~\eqref{eq:updateRule} and the experimental parameters given in Sec.~\ref{sec:experimentalSetup}. The discontinuities in $\lambda_t$ indicate the detection of a photon. \label{fig:fig4}}
\end{figure}

To acquire fluorescence trajectories, $594$-$\textrm{nm}$ excitation is applied continuously and photon counts are recorded for $30\,\textrm{s}$ in bins of $100\,\mu\textrm{s}$, with a small separation of $8.3\,\textrm{ns}$ necessary for our field-programmable gate array (FPGA) card to process the counts. These time scales are chosen such that the total duration ($30\,\textrm{s}$) is much greater than the time scale for ionization and recombination, and each time bin ($100\,\mu\textrm{s}$) is much shorter. A subset of such a time trace is illustrated in Fig.~\ref{fig:fig4}(a) with data rebinned to $10\,\textrm{ms}$. The experiment is repeated to obtain statistics on fluorescence and ionization rates, interleaved with a tracking step ($532$-$\textrm{nm}$ green laser) that ensures the sample does not drift relative to the focal point of the microscope. The experimental sequence is illustrated in Fig.~\ref{fig:fig3}(c). Such data were acquired for excitation powers ranging from $0.55\,\mu\textrm{W}$ to $5\,\mu\textrm{W}$ (as measured going into the objective). For comparison, the saturation intensity at $594\,\textrm{nm}$ is estimated to be $2.5\,\textrm{mW}$. We verified that $\gamma_+$ and $\gamma_-$ scale as the laser intensity and that $\Gamma_+$ and $\Gamma_-$ scale as the laser intensity squared, as shown in Ref.~\cite{aslam2013}. Data sets were measured for two $\left<111\right>$ oriented NV centers, which yielded similar results. Very weak excitation is used to ensure that $\gamma_+$ and $\gamma_-$ are much larger than $\Gamma_+$ and $\Gamma_-$.

For definiteness, we restrict the following analysis to a single power setting. Specifically, the data sets for the lowest laser power setting of $0.55\,\mu\textrm{W}$ were used to extract the parameters $\gamma_{\pm}$ and $\Gamma_{\pm}$, as detailed in Appendix~\ref{app:ratesExtraction}. We obtain the rates $\gamma_+ \approx 720\,\textrm{Hz}$, $\gamma_- \approx 50\,\textrm{Hz}$, $\Gamma_+ \approx 3.6\,\textrm{Hz}$ and $\Gamma_- \approx 0.98\,\textrm{Hz}$, which are required to construct an algorithm for an adaptive decision rule.

\subsection{Readout error analysis \label{sec:errorRate}}

In Refs.~\cite{aslam2013,shields2015}, single-shot readout of the charge state was performed by counting the total number of photons detected in the interval $\left[0,t_f\right]$ (see Appendix~\ref{app:photonCounting}). In contrast, implementing the adaptive decision rule requires an efficient algorithm to update the likelihood ratio after each time bin~\cite{gambetta2007,myerson2008,gehr2010,danjou2014,ng2014,gammelmark2014,wolk2015}. A hidden-Markov-model algorithm suitable to the NV-center charge readout has been developed in, e.g., Refs.~\cite{ng2014,wolk2015}. Here, we use an equivalent quantum trajectory approach, which is more easily generalized to quantum systems with multiple levels and coherent internal dynamics. Moreover, we find an update rule that is valid for an arbitrary value of the bin size $\delta t$. In particular, more than one ionization or recombination event may occur during one time bin, an occurrence that becomes more likely at high illumination power~\cite{aslam2013}. The update rule of Ref.~\cite{wolk2015} is recovered when $\Gamma_{\pm} \delta t \ll 1$.

As shown in Appendix~\ref{app:quantumTrajectoryApproach}, the likelihood ratio $\Lambda_N = P(\psi_N|+)/P(\psi_N|-)$ after $N$ bins is given by
\begin{align}
	\Lambda_N = \frac{\textrm{Tr}\left[\mathbf{M}(\delta n_{N-1})\,\dots\,\mathbf{M}(\delta n_0)\, \boldsymbol{\ell}_0^{(+)} \right]}{\textrm{Tr}\left[\mathbf{M}(\delta n_{N-1})\,\dots\,\mathbf{M}(\delta n_0)\, \boldsymbol{\ell}_0^{(-)} \right]}. \label{eq:updateRule}
\end{align}
In Eq.~\eqref{eq:updateRule}, $\boldsymbol{\ell}_0^{(+)} = (1,0)^T$ and $\boldsymbol{\ell}_0^{(-)} = (0,1)^T$ are initial state vectors representing $\ket{+}$ and $\ket{-}$, respectively. Here, the trace $\textrm{Tr}$ of a vector is defined as the sum of its elements. The matrix $\mathbf{M}(\delta n)$ is a $2\times 2$ update matrix for the detection of $\delta n$ photons in a bin. The exact form of the update matrices is given in Appendix~\ref{app:quantumTrajectoryApproach}. They can be stored for all $\delta n \leq \delta n_{\textrm{max}}$, where $\delta n_{\textrm{max}}$ is the maximum number of photons with a non-negligible probability to occur. The likelihood ratio, Eq.~\eqref{eq:updateRule}, can then be updated in real time through simple matrix multiplication. An example of the application of Eq.~\eqref{eq:updateRule} to experimental data is given in Fig.~\ref{fig:fig4}(b). Note that $\mathbf{M}$ is nothing but the measurement superoperator for a classical Markov process subject to Poissonian noise. The approach can thus be generalized to any Markovian quantum trajectory by substituting $\mathbf{M}$ with the appropriate measurement superoperator for direct detection and the vectors $\boldsymbol{\ell}_0^{(\pm)}$ by any pair of initial states to be discriminated. In particular, the standard spin detection of the NV center modeled in, e.g., Refs.~\cite{manson2006,robledo2011-2,doherty2013} can be processed in this way.

 In the following, we first use Monte Carlo simulations to assess the ideal theoretical performance of the adaptive decision rule. We then apply the adaptive decision rule to experimental data to demonstrate a speedup under real experimental conditions. In all cases, we choose the experimental time bin $\delta t = 0.1 \,\textrm{ms}$ and a maximum acquisition time $t_M = N \delta t = 25\,\textrm{ms}$. We use this value of $\delta t$ and the rates given in Sec.~\ref{sec:experimentalSetup} to calculate and store the update matrices $\mathbf{M}(\delta n)$ for $\delta n=0,1,\ldots,\delta n_{\textrm{max}}$. Here, $\delta n_{\textrm{max}}=5$ is chosen to match the maximum number of photons observed in any one time bin of the experimental data sets. The residual probability of observing more than $\delta n_{\textrm{max}}$ photons in one time bin is $\approx 10^{-10}$.

\subsubsection{Monte Carlo simulations \label{sec:simulations}}

\begin{figure}
\centering
\includegraphics[width=\columnwidth]{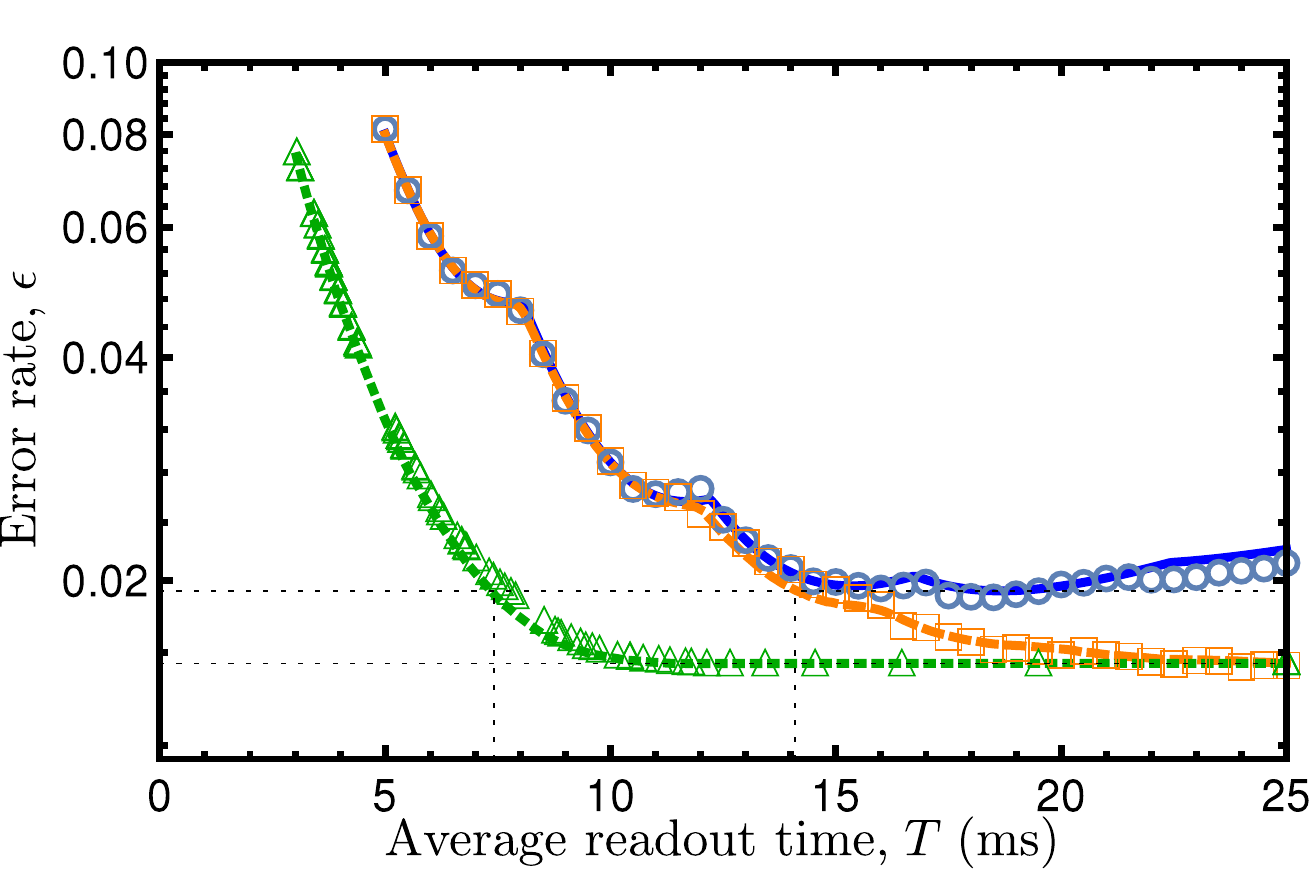}
\centering
\caption{Error rate as a function of average readout time for the experimental parameters of Sec.~\ref{sec:experimentalSetup}. The lines give the Monte Carlo simulated error rate, and the symbols give the experimental values corrected for preparation errors using Eq.~\eqref{eq:preparationCorrection}. We display the error rate for the photon-counting model discussed in Appendix~\ref{app:photonCounting} (solid blue and blue circles), for the nonadaptive MLE (dashed orange and orange squares), and for the adaptive MLE (dotted green and green triangles). The horizontal dotted black lines indicate the minimum error rate of the photon-counting (1.9\%) and MLE methods (1.5\%). As indicated by the vertical dotted black lines, the speedup of the adaptive MLE compared to the nonadaptive MLE is $t_f/T \approx 1.9$ when the target error rate is the minimum error rate of the photon-counting method. \label{fig:fig5}}
\end{figure}

For both initial states $\ket{\pm}$, we use the procedure described in Appendix~\ref{app:monteCarlo} to generate $1\times 10^{6}$ random trajectories $\psi_N$ in the interval $\left[0, N \delta t\right]$. We then apply both the nonadaptive and the adaptive MLE to each trajectory to obtain a Monte Carlo estimate of the average readout time $T$ and error rate $\epsilon$, as detailed in Appendix~\ref{app:monteCarlo}. The resulting error-rate curves are shown in Fig.~\ref{fig:fig5}. For comparison, we also show the error rate of the photon-counting method modeled in Ref.~\cite{shields2015} and summarized in Appendix~\ref{app:photonCounting}. In this approach, the total number of detected photons is compared to a threshold to choose the state. As expected, the nonadaptive and adaptive MLE achieve a somewhat lower minimum error rate than photon counting. More importantly, however, we see that the adaptive MLE achieves the minimum error rate of the photon-counting method in approximately half the time necessary for the nonadaptive MLE to achieve the same goal. Thus, the bandwidth of the already efficient NV-center charge readout is doubled through signal processing alone. We have verified that a similar advantage also exists for the experimental parameters of Ref.~\cite{shields2015}. Since this speedup enables a larger number of NV-center charge measurements to be performed in a given amount of time, it should lead to a substantial improvement in the sensitivity of the spin-to-charge conversion magnetometer in cases where the measurement duty cycle is limited by readout time.

We note that the parameters given in Sec.~\ref{sec:experimentalSetup} satisfy the condition for which the NV-center charge readout reduces to the state-dependent decay readout of Sec.~\ref{sec:stateDependentDecay}, $\tau^{-1} = \gamma_+ > \max(\gamma_-,\Gamma_+,\Gamma_-)=\gamma_-$. It is therefore plausible that the speedup $\approx 2$ obtained in Fig.~\ref{fig:fig5} has at least partially the same origin as that given in Eq.~\eqref{eq:decaySpeedup}. Indeed, the speedup $\approx 2$ observed in Ref.~\cite{noek2013} was essentially explained by the state-dependent decay model for a readout whose dynamics are formally the same as the NV-center charge readout dynamics. However, the model of Sec.~\ref{sec:stateDependentDecay} fails when the maximum readout time $t_M$ becomes comparable to $\gamma_-^{-1}\approx 20\,\textrm{ms}$, resulting in a much higher saturation error rate in Fig.~\ref{fig:fig5} than that predicted by Eq.~\eqref{eq:errorEmission} with $\tau = \gamma_+^{-1} \approx 1.4\,\textrm{ms}$ and $t_f = t_M = 25\,\textrm{ms}$. In our experiment, it is therefore necessary to use the update rule of Eq.~\eqref{eq:updateRule} to obtain an accurate estimate of the error rate. In a context where it may be favorable or necessary to move towards the regime of the high-fidelity Gaussian latching readout of Sec.~\ref{sec:gaussianLatchingReadout}, speedups larger than $2$ should be possible. Entering this regime may not be advantageous for the NV-center charge readout since the increase in excitation intensity required to reach the Gaussian regime results in a detrimental increase in the ionization and recombination rates. However, such an advantage may be possible in other systems with similar dynamics.

\subsubsection{Experimental verification \label{sec:modelVerification}}

In the simulations of Sec.~\ref{sec:simulations}, the statistics of the simulated fluorescence trajectories are perfectly described by the two-state model assumed for readout. In real experimental conditions, however, the charge dynamics may deviate from this model. In this section, we show that the adaptive-decision speedup is robust to imperfections in our modeling by applying the adaptive decision rule to experimental data. To independently verify our model of charge dynamics, we split our data sets into a calibration set and a testing set. The calibration set is used solely to extract the experimental values of the rates $\gamma_\pm$ and $\Gamma_\pm$ given in Sec.~\ref{sec:experimentalSetup}. The extracted parameters and Eq.~\eqref{eq:updateRule} are then used on the testing set to verify the model. The details are given in Appendix~\ref{app:modelVerification}.

We first perform a preliminary verification by comparing the experimental distribution of log-likelihood ratios to the distribution predicted by Monte Carlo simulations [see Appendix~\ref{app:modelVerification}]. We find a close agreement between experiment and theory, showing that the model of charge dynamics discussed in Sec.~\ref{sec:chargeDynamics} provides a good description of the statistics of the experimental trajectories. We can therefore use our model to perform approximate preparation of the charge state in postselection. We use this preparation to verify that the adaptive-decision speedup exists when the adaptive decision rule is applied to experimental data. From the testing set, we prepare $8307$ ($30693$) trajectories of $25\,\textrm{ms}$ with initial states $\ket{+}$ ($\ket{-}$). We then read out the state for each trajectory using the photon-counting, nonadaptive MLE, and adaptive MLE methods (i.e. assuming equal prior probabilities). Comparing the result of the readout to the preparation allows us to estimate the experimental error rate $\tilde{\epsilon} = (\tilde{\epsilon}_+ + \tilde{\epsilon}_-)/2$, where $\tilde{\epsilon}_\pm$ are the error rates conditioned on the preparation in state $\ket{\pm}$. To better compare the experimental results with theory, we then fit the experimental error-rate curves to the theoretical prediction, accounting for an additional preparation error $\eta$ [see Eq.~\eqref{eq:preparationCorrection}]. This procedure gives $\eta=2.22\%$ as a single fit parameter. We emphasize that this adjustment is a uniform transformation of the error rate for all times and hence does not affect the measured speedup. The experimental error rate $\epsilon$ obtained after compensating for preparation error is shown along with the error rate determined by Monte Carlo simulation in Fig.~\ref{fig:fig5} [the uncompensated error rate $\tilde{\epsilon}$ is shown in Appendix~\ref{app:modelVerification} for comparison]. The experimentally measured error rate thus fits the theoretical prediction very well. Therefore, we conclude that the adaptive-decision speedup $\approx 2$ discussed in Sec.~\ref{sec:simulations} persists when the adaptive decision rule is applied to real experimental data in spite of possible systematic errors and imperfect modeling.

\subsubsection{Unequal prior probabilities \label{sec:errorsBias}}

A balanced probability distribution for the initial state, $P(+)=P(-)$, is often desirable because it enables the extraction of one bit of information per measurement. However, in applications such as the magnetometry protocol of Ref.~\cite{shields2015}, the probability distribution of the initial state is, in reality, unbalanced, $P(+)\neq P(-)$. In general, the dependence of the adaptive-decision speedup on the prior probabilities $P(\pm)$ is nontrivial and may depend on the particularities of the readout dynamics. A general analysis lies beyond the scope of this work. Here, we nevertheless discuss how to account for unequal prior probabilities. Moreover, we show that the adaptive-decision speedup of the experimentally relevant NV-center charge readout discussed in Sec.~\ref{sec:simulations} persists even for unequal prior probabilities. We distinguish the case where $P(\pm)$ are unknown from the case where $P(\pm)$ are known.

\begin{figure}
\centering
\includegraphics[width=\columnwidth]{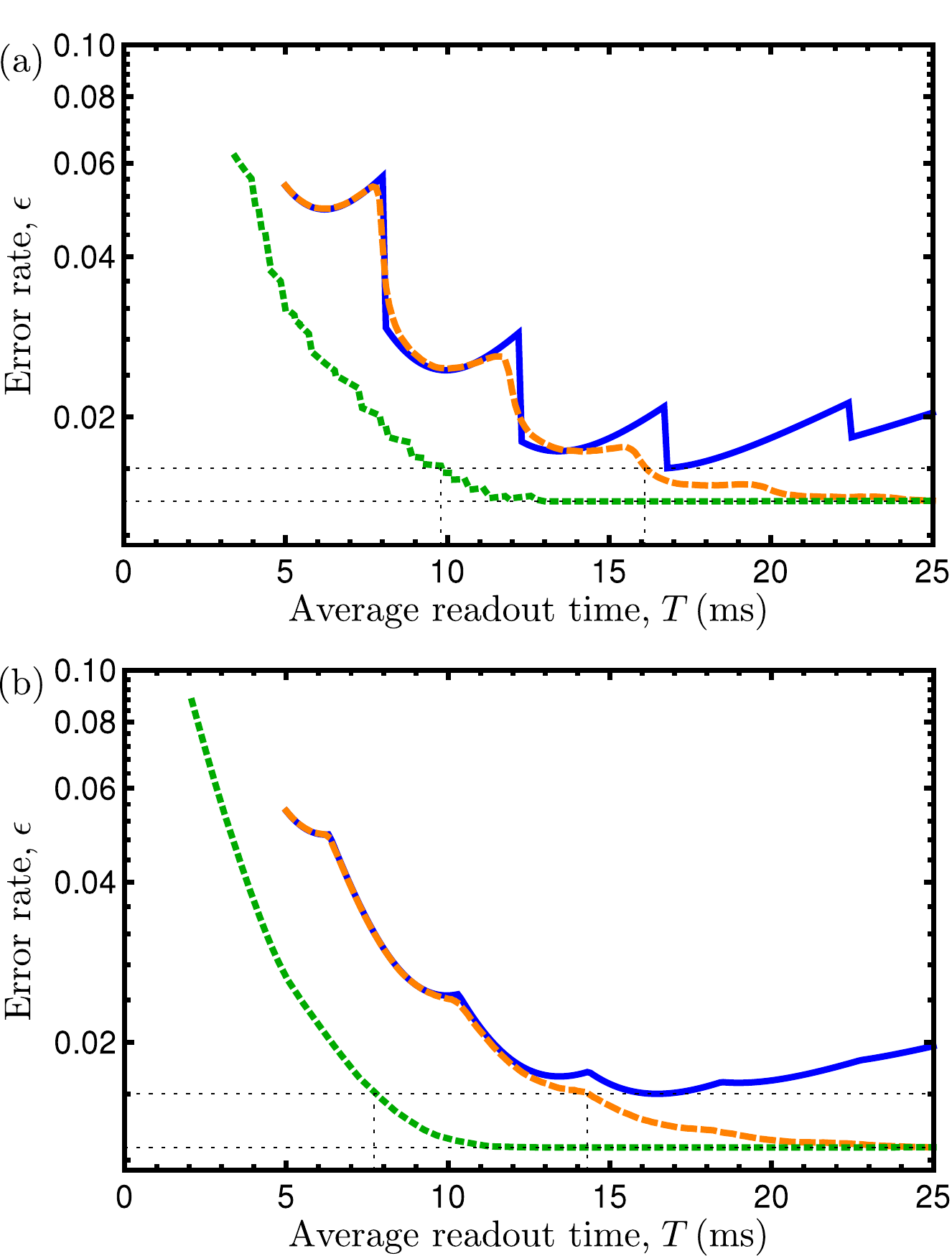}
\centering
\caption{Monte Carlo simulated error rate for an unbalanced initial charge distribution with $P(+) = 0.25$ when (a) the value of $P(+)$ is unknown (MLE) and (b) the value of $P(+)$ is known (MAP). In both cases, we show the error rate for the photon-counting method of Ref.~\cite{shields2015} (solid blue), the nonadaptive readout (dashed orange) and the adaptive readout (dotted green). The horizontal dotted black lines indicate the minimum error rates of the photon-counting method ($1.6\%$ in both cases) and of the optimal methods ($1.4\%$ for the MLE and $1.3\%$ for the MAP). The dotted vertical lines show that significant adaptive-decision speedups of $t_f/T \approx 1.6$ and $t_f/T \approx 1.8$ are obtained when using the MLE and MAP, respectively. The discontinuous features in (a) occur because the MLE uses a nonoptimal discrimination threshold when $P(\pm)\neq 0.5$. \label{fig:fig6}}
\end{figure}

When the prior probabilities are unknown, the best strategy is to calibrate the readout thresholds $\lambda_{\textrm{th}}$ and $\lambda_{\pm}$ by assuming that the prior probabilities are equal. This leads to the MLE readout discussed in Secs.~\ref{sec:simulations} and ~\ref{sec:modelVerification}. Under this assumption, we obtain the error rates $\epsilon^{(\pm)}$ and average times $T^{(\pm)}$ conditioned on the state $\ket{\pm}$. When the MLE readout is used on an unbalanced sample of charge states, the error rate and average readout times are then given by $\epsilon = P(+)\epsilon^{(+)} + P(-)\epsilon^{(-)}$ and $T = P(+)T^{(+)} + P(-)T^{(-)}$, respectively. The simulated error rate as a function of average readout time in this scenario is shown in Fig.~\ref{fig:fig6}(a) for the experimental parameters of Sec.~\ref{sec:experimentalSetup} and $P(+) = 0.25$. We also show the corresponding error-rate curve for the photon-counting method (see Appendix~\ref{app:photonCounting}). Using the minimum error rate of the photon-counting method as a reference, we find that a substantial speedup $t_f/T \approx 1.6$ persists even when the prior probabilities are unknown.

When the prior probabilities are known, they can be used to improve the MLE readout by adjusting the thresholds to directly minimize $\epsilon = P(+)\epsilon^{(+)} + P(-)\epsilon^{(-)}$ for constant $T = P(+)T^{(+)} + P(-)T^{(-)}$. This leads to the MAP readout mentioned in Sec.~\ref{sec:adaptiveDecisionRule} (see Appendix~\ref{app:monteCarlo} for details). We note that in a given application, the prior probabilities can easily be determined by using the MLE to estimate the relative proportion of $\ket{+}$ and $\ket{-}$. The simulated error rate as a function of average readout time when using the MAP is shown in Fig.~\ref{fig:fig6}(b) for the same experimental parameters and $P(+)=0.25$. We find an adaptive-decision speedup $t_f/T \approx 1.8$ using the minimum error rate of the photon-counting method as a reference, similar to the value obtained in Sec.~\ref{sec:simulations} for $P(+)=0.5$. Therefore, we conclude that the adaptive-decision speedup persists for significant deviations from a balanced initial-state distribution.

\section{Parametric improvement in speedup \label{sec:parametricSpeedup}}

From the examples of Secs.~\ref{sec:boundedSpeedup} and \ref{sec:chargeReadoutNV}, we might be tempted to conclude that the speedup is fundamentally bounded by a constant of order unity. Here, we show that there exist readout schemes where the speedup can become arbitrarily large in the limit of low error rate. 

\begin{figure}
\centering
\includegraphics[width=\columnwidth]{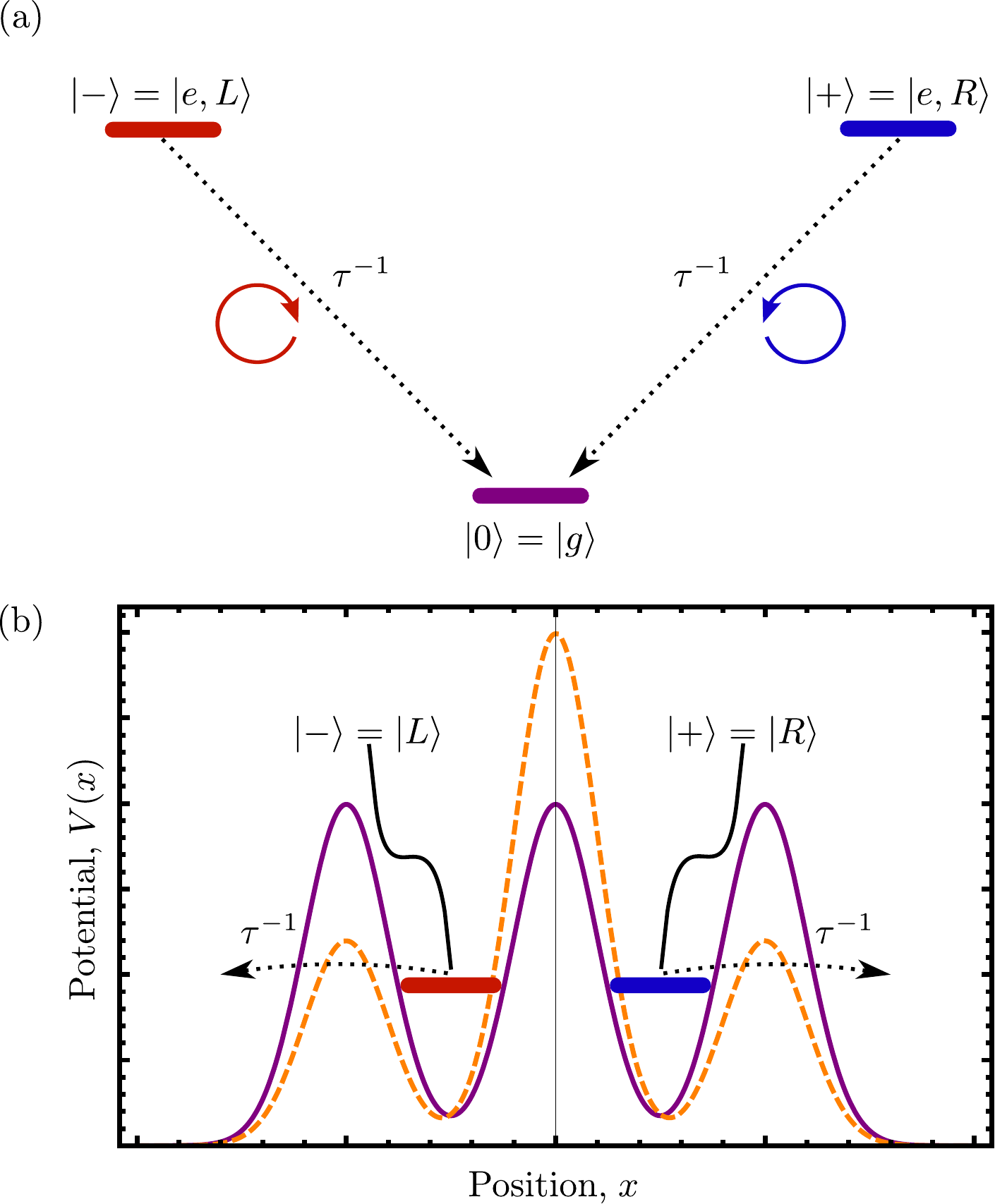}
\centering
\caption{(a) Readout scheme based on state-dependent polarization of an emitted photon. When the state is $\ket{+} = \ket{e,R}$ ($\ket{-} = \ket{e,L}$) in the excited subspace, a right (left) circularly polarized photon is emitted. Both states decay to the ground state $\ket{0} = \ket{g}$ on a time scale $\tau$. (b) Similar scheme for readout of the charge state of a double quantum dot. When the electron is in the right (left) island $\ket{R}$ ($\ket{L}$), it is detected after tunneling into the right (left) lead. Both states tunnel into the leads on a time scale $\tau$, leaving the double quantum dot in the empty state $\ket{0}$. \label{fig:fig7}}
\end{figure}

To show this, we consider a simple variation of the state-dependent decay readout analyzed in Sec.~\ref{sec:stateDependentDecay}. We now suppose that the states $\ket{+}$ and $\ket{-}$ both decay to a third state $\ket{0}$ through different decay channels $R$ and $L$, respectively. The two channels may be associated with, e.g., different polarizations of emitted photons or two different leads into which the electron of a double-quantum-dot charge qubit can tunnel (see Fig.~\ref{fig:fig7}). If the two decay channels can be discriminated, e.g., with the help of a polarization analyzer, the state $\ket{\pm}$ can be read out. For simplicity we assume that both states decay with the same probability per unit time $\tau^{-1}$. 

When using a nonadaptive decision rule, we wait a time $t_f \gtrsim \tau$ and base the decision on whether a $R$ or $L$ decay event has been detected. If no event is detected, we choose the state at random with equal probability. Assuming, for simplicity, that the channels $R$ and $L$ can be perfectly discriminated, the error rate is given by the probability that no decay event has occurred up to time $t_f$, $e^{-t_f/\tau}$, multiplied by the probability $1/2$ that the random decision fails. Thus, the error rate is still given by Eq.~\eqref{eq:errorEmission}.

The adaptive decision rule, in contrast, stops the readout as soon as either a decay event is detected or a maximum time $t_M$ is reached. Following the same argument as for the nonadaptive decision rule, the error rate is $\epsilon=e^{-t_M/\tau}/2$. To achieve the same error rate as the nonadaptive decision rule, we must therefore choose $t_M = t_f$. However, we find that the average readout time is now $T = (1-e^{-t_f/\tau})\tau$. As the error rate decreases, $t_f \rightarrow \infty$, the average readout time tends to a constant $\tau$ so that
\begin{align}
	\frac{t_f}{T} \sim \ln\left(\frac{1}{2\epsilon}\right) \rightarrow \infty, \;\;\; (\epsilon\rightarrow 0). \label{eq:parametricSpeedup}
\end{align}
Therefore, the speedup increases without bound as the error rate $\epsilon$ is decreased. This parametric improvement in speedup is achieved by transferring all the information about the state to the channel degree of freedom, $R$ or $L$. This means that the readout no longer relies on discriminating decay from the absence of decay. We can thus arrange the readout so that both $\ket{-}$ and $\ket{+}$ decay on a time scale $\tau$. Therefore, it is no longer necessary to wait for the entire duration $t_f$ when the state is $\ket{-}$.

Such a scheme could be applied in a variety of scenarios. For example, suppose that two atomic excited states, $\ket{+} = \ket{e,R}$ and $\ket{-} = \ket{e,L}$, both decay to the ground state $\ket{0} = \ket{g}$ by emitting photons with right and left circular polarizations, respectively. This situation is depicted in Fig.~\ref{fig:fig7}(a). A polarization analyzer could then identify both states in a time $\approx \tau$, leading to the parametric improvement of Eq.~\eqref{eq:parametricSpeedup}. In another example, suppose that we want to discriminate between an electron being in the rightmost ($\ket{+} = \ket{R}$) and leftmost ($\ket{-} = \ket{L}$) islands of a double quantum dot. We imagine that, as shown in Fig.~\ref{fig:fig7}(b), the potential barriers between the dots and leads are lowered at the beginning of the readout phase. An electron in the right (left) dot will then tunnel out into the right (left) lead and be detected, with the left and right detectors playing the role of the polarization analyzers in the previous example. In perfect analogy with Fig.~\ref{fig:fig7}(a), stopping readout as soon as an electron is detected in either lead instead of waiting for a fixed time $t_f$ then leads to a parametric improvement in speedup, Eq.~\eqref{eq:parametricSpeedup}. Note that while the assumption of perfect detection efficiency is difficult to realize for photon detection, it is usually not a limitation for the detection of electron charges in quantum dots.

\section{Conclusion \label{sec:conclusion}}

In summary, we have established the maximum achievable ``adaptive-decision speedups'' for several physical readout models in the high-fidelity limit. To achieve this, we formulated the adaptive-decision problem in terms of a first-passage time problem. We obtained bounds for the speedup of two commonly encountered readout models. Specifically, we have shown that the adaptive decision rule can speed up the Gaussian latching readout (Sec.~\ref{sec:gaussianLatchingReadout}) by up to a factor of $4$ in the limit of high readout fidelity, while it can speed up readouts relying on state-dependent decay (Sec.~\ref{sec:stateDependentDecay}) by up to a factor of $2$. To study the achievable speedup in a real-world scenario, we applied the adaptive decision rule to a readout of the charge state of a NV center in diamond using a quantum trajectory formalism that incorporates the NV-center charge dynamics and accounts for experimental imperfections such as dark counts and imperfect collection efficiency. Although the NV-center charge readout reduces to the two aforementioned models in distinct limits, its dynamics are in an intermediate regime for typical experimental parameters. We have shown that a significant speedup can be achieved under these conditions. Specifically, we found a speedup $\approx 2$ both in our experiment and using the experimental parameters of Ref.~\cite{shields2015}. Finally, we have proposed a readout scheme that leads to an unbounded speedup as the fidelity is increased, in stark contrast to the common readout models discussed previously. This scheme relies on discriminating between two distinct decay channels instead of discriminating between decay and absence of decay. We have further provided avenues to realize such a scheme in atomic or quantum-dot systems.

Our results are already applicable to a wide variety of systems, and they provide a direction for the optimization of measurement bandwidth in experimentally relevant readouts of quantum states. In particular, we have shown that magnetometry based on the NV-center spin-to-charge-conversion readout can be improved with currently achievable experimental parameters. Moreover, our results show that apparent limitations to the adaptive speedup can be overcome by careful redesign of the readout dynamics. In this work, our goal was to analyze the fundamental limitations of the adaptive-decision speedup for extreme limits of several physical readout schemes. We expect direct extensions of the first-passage time formalism developed here to allow for the derivation of analytical speedup bounds in the presence of, e.g., unwanted state relaxation, dark counts, or imperfect photon collection efficiency. On a more fundamental level, our analysis provides the necessary framework to study how the addition of measurement feedback and coherent readout dynamics can modify the maximal speedups discussed here. Moreover, this formalism may allow for the direct characterization of the achievable speedups for parameter estimation, or multiple-state discrimination in general, subject to experimentally relevant noise.

\section*{Acknowledgments}

We acknowledge financial support from NSERC, CIFAR, INTRIQ, and the Canada Research Chairs Program.


\appendix

\section{Analytical treatment of the adaptive Gaussian latching readout \label{app:gaussianAnalyticalResults}}

In this section, we sketch the derivation of the error rate and average readout time for both the nonadaptive and the adaptive decision rule applied to the Gaussian latching readout. According to Eq.~\eqref{eq:signalDistribution}, the expectation and covariance of the trajectory $\psi_t(t')$ conditioned on the state $\ket{\pm}$ are
\begin{align}
\begin{split}
	&\mathbb{E}\left[\psi_t(t')|\pm\right] = \pm 1, \\
	&\textrm{Cov}\left[\psi_t(t'),\psi_t(t'')|\pm\right] = r^{-1} \delta(t'-t''), \label{eq:moments}
\end{split}
\end{align}
for $t' < t$ and $t'' < t$. Because the noise is Gaussian, all higher cumulants vanish. Correspondingly, the expectation and variance of the log-likelihood ratio $\lambda_t = 2 r \int_0^t dt'\, \psi_t(t')$ are $\mathbb{E}(\lambda_t|\pm) = \pm 2 r t$ and $\textrm{Var}(\lambda_t|\pm) = 4 r t$. Therefore, $\lambda_t$ is a simple drift-diffusion process distributed according to
\begin{align}
\begin{split}
	P(\lambda_t|\pm) &\equiv G^{(\pm)}(\lambda_t,t) \\
	&= \frac{1}{\sqrt{8\pi r\,t}} \exp{\left[-\frac{(\lambda_t \mp 2 r t)^2}{8 r t}\right]}. \label{eq:driftDiffusionGreensFunction}
	\end{split}
\end{align}
Here, $G^{(\pm)}(\lambda,t)$ is the Green's function that solves the associated drift-diffusion equation for $\lambda_t$.

For a fixed readout time $t_f$, the state is chosen according to the sign of $\lambda_{t_f}$. The error rates conditioned on the initial state being $\ket{\pm}$, $\epsilon^{(\pm)}$, are thus
\begin{align}
\begin{split}
	&\epsilon^{(+)} = \int_{-\infty}^0 d\lambda_{t_f} \,P(\lambda_{t_f}|+), \\
	&\epsilon^{(-)} = \int_0^{\infty} d\lambda_{t_f}\,P(\lambda_{t_f}|-).
\end{split}
\end{align}
By symmetry of the problem, $\epsilon^{(+)} = \epsilon^{(-)}$. Using Eq.~\eqref{eq:driftDiffusionGreensFunction}, we can then calculate the error rate $\epsilon = \left[\epsilon^{(+)} + \epsilon^{(-)}\right]/2 = \epsilon^{(\pm)}$:
\begin{align}
	\epsilon = \frac{1}{2} \textrm{erfc}\left(\sqrt{\frac{r\,t_f}{2}}\right).
\end{align}

To assess the performance of the adaptive decision rule, we must calculate both the error rate $\epsilon = \left[\epsilon^{(+)} + \epsilon^{(-)}\right]/2$ and the average readout time $T = \left[T^{(+)} + T^{(-)}\right]/2$, where $T^{(\pm)}$ is the average readout time conditioned on the state being $\ket{\pm}$. Using the symmetry of the problem again, we have $\epsilon = \epsilon^{(\pm)}$ and $T = T^{(\pm)}$. We may therefore assume that the state is $\ket{+}$ without loss of generality. We now reformulate the calculation of $\epsilon$ and $T$ as a first-passage time problem~\cite{risken1989,klebaner2005} (an alternative formalism is given in Ref.~\cite{tantaratana1982}). More precisely, let $\varepsilon(\lambda,t)$ and $\mathcal{T}(\lambda,t)$ be the error rate and average readout time conditioned on knowing that the log-likelihood ratio is $\lambda$ at time $t$. We wish to obtain $\epsilon = \varepsilon(0,0)$ and $T = \mathcal{T}(0,0)$. Because drift diffusion is a Markov process, we can use Eq.~\eqref{eq:driftDiffusionGreensFunction} to condition the values of $\varepsilon$ and $\mathcal{T}$ on their possible values at time $t+\delta t$:
\begin{align}
\begin{split}
	&\varepsilon(\lambda,t) = \int_{-\infty}^{\infty} \!dx\, G^{(+)}(x,\delta t) \varepsilon(\lambda+x,t +\delta t),\\
	&\mathcal{T}(\lambda,t) = \delta t + \int_{-\infty}^{\infty} \!dx\, G^{(+)}(x,\delta t) \mathcal{T}(\lambda+x,t+\delta t). \label{eq:conditioning}
\end{split}
\end{align}
We then expand $\varepsilon$ and $\mathcal{T}$ around $x=0$ and $\delta t = 0$ on both sides and keep all terms of order $\delta t$. In the limit $\delta t \rightarrow 0$, Eqs.~\eqref{eq:conditioning} then reduce to Kolmogorov backward partial differential equations~\cite{risken1989,klebaner2005} of the drift-diffusion type~\footnote{We note that backward equations can also be obtained for the case of Poissonian noise. In this case, each partial differential equation is replaced by a discrete set of coupled rate equations. See, for example, Ref.~\cite{feller1968}.}:
\begin{align}
\begin{split}
	-\frac{1}{2r}\frac{\del \varepsilon(\lambda,t)}{\del t} &= \frac{\del \varepsilon(\lambda,t)}{\del \lambda} + \frac{\del^2 \varepsilon(\lambda,t)}{\del \lambda^2}, \\
	-\frac{1}{2r}\frac{\del \mathcal{T}(\lambda,t)}{\del t} &= \frac{\del \mathcal{T}(\lambda,t)}{\del \lambda} + \frac{\del^2 \mathcal{T}(\lambda,t)}{\del \lambda^2} + \frac{1}{2r}. \label{eq:driftDiffusionPDE}
\end{split}
\end{align}
The adaptive decision rule is implemented by setting appropriate boundary conditions. Since we have assumed that the state is $\ket{+}$, an error occurs only when we stop with $\lambda_t < 0$. We must therefore have $\varepsilon(\pm \bar{\lambda},t) = \theta(\mp\bar{\lambda})$ and $\varepsilon(\lambda,t_M) = \theta(-\lambda)$. Similarly, the remaining readout time after stopping must vanish, $\mathcal{T}(\pm \bar{\lambda},t) = 0$ and $\mathcal{T}(\lambda,t_M) = 0$. Here, $\theta(x)$ is the Heaviside step function.

To solve Eqs.~\eqref{eq:driftDiffusionPDE} analytically, we first take $t_M\rightarrow \infty$. The corresponding solutions $\varepsilon_\infty(\lambda)$ and $\mathcal{T}_\infty(\lambda)$ are independent of $t$ and satisfy the ordinary differential equations:
\begin{align}
\begin{split}
		&\frac{\del \varepsilon_\infty(\lambda)}{\del \lambda} + \frac{\del^2 \varepsilon_\infty(\lambda)}{\del \lambda^2} = 0, \\
		&\frac{\del \mathcal{T}_\infty(\lambda)}{\del \lambda} + \frac{\del^2 \mathcal{T}_\infty(\lambda)}{\del \lambda^2} + \frac{1}{2r} = 0,
\end{split} \label{eq:stationaryPDE}
\end{align}
subject to the boundary conditions $\epsilon_\infty(\pm \bar{\lambda}) = \theta(\mp \bar{\lambda})$ and $\mathcal{T}_\infty(\pm \bar{\lambda}) = 0$. Solving Eqs.~\eqref{eq:stationaryPDE} gives Eq.~\eqref{eq:errorAdaptive}:
\begin{align}
\begin{split}
&\epsilon_\infty = \varepsilon_\infty(0) = \frac{1}{1+e^{\bar{\lambda}}} \\
&T_\infty = \mathcal{T}_\infty(0) = \frac{\bar{\lambda}}{2r}\tanh\left(\frac{\bar{\lambda}}{2}\right). \label{eq:errorAdaptiveApp}
\end{split}
\end{align}
To obtain the solution for finite $t_M$, we write $\varepsilon(\lambda,t) = \varepsilon_\infty(\lambda) + \eta(\lambda,t)$ and $\mathcal{T}(\lambda,t) = \mathcal{T}_\infty(\lambda) + \zeta(\lambda,t)$. Substituting these expressions in Eqs.~\eqref{eq:driftDiffusionPDE}, we find that $\eta$ and $\zeta$ satisfy the homogeneous equations
\begin{align}
	\begin{split}
	-\frac{1}{2r}\frac{\del \eta(\lambda,t)}{\del t} &= \frac{\del \eta(\lambda,t)}{\del \lambda} + \frac{\del^2 \eta(\lambda,t)}{\del \lambda^2}, \\
	-\frac{1}{2r}\frac{\del \zeta(\lambda,t)}{\del t} &= \frac{\del \zeta(\lambda,t)}{\del \lambda} + \frac{\del^2 \zeta(\lambda,t)}{\del \lambda^2}, \label{eq:transientEquations} 
\end{split}
\end{align}
subject to the boundary conditions $\eta(\pm \bar{\lambda},t) = 0$, $\eta(\lambda,t_M) = \theta(-\lambda)-\varepsilon_\infty(\lambda)$ and $\zeta(\pm \bar{\lambda},t) = 0$, $\zeta(\lambda,t_M) = -\mathcal{T}_\infty(\lambda)$, respectively. For each of the Eqs.~\eqref{eq:transientEquations}, we decompose the boundary condition at time $t_M$ in terms of the basis of right eigenfunctions of the drift-diffusion operator $\partial_\lambda + \partial_\lambda^2$. The eigenfunctions are chosen to vanish for $\lambda = \pm \bar{\lambda}$. We then propagate each component backwards in time to find $\eta(0,0)$ and $\zeta(0,0)$. This gives the exact analytical expressions:
\begin{align}
\begin{split}
	&\epsilon = \epsilon_{\infty} + \sum_{m=0}^{\infty} A_m e^{-\alpha_m r\,t_M}, \\
	&r\,T = r\,T_{\infty} + \sum_{m=0}^{\infty} B_m e^{-\alpha_m r\,t_M}. \label{eq:errorAdaptiveTruncated}
\end{split}	
\end{align}
Here, we define
\begin{align}
\begin{split}
	&A_m = \frac{2 \bar{\lambda}}{4 \pi^2 (m+1/2)^2 + \bar{\lambda}^2}, \\
	&B_m = -\frac{16 \pi \bar{\lambda}^2 \cosh{(\frac{\bar{\lambda}}{2})} (-1)^m (m+1/2)}{\left[4\pi^2 (m+1/2)^2 + \bar{\lambda}^2\right]^2}, \\
	&\alpha_m = \frac{1}{2} + \frac{2 \pi^2 (m+1/2)^2}{\bar{\lambda}^2}. 
\end{split}
\end{align}
Equation~\eqref{eq:errorAdaptiveTruncated} reduces to Eq.~\eqref{eq:errorAdaptiveApp} when $r\,t_M \gg \min(1,\bar{\lambda}^2)$.

\section{Derivation of the update matrices \label{app:quantumTrajectoryApproach}}

In this section, we derive the form of the update matrices $\mathbf{M}(n)$ using a number-resolved quantum trajectory formalism~\cite{cook1981,emary2007}. The NV-center charge state at any given time $t$ can be described by a state vector $\boldsymbol{\rho} = (\rho_+, \rho_-)^T$, where $\rho_{\pm}$ is the probability of finding the charge state $\ket{\pm}$. The state vector obeys the equation of motion of a Markovian two-level fluctuator:
\begin{align}
	\dot{\boldsymbol{\rho}}(t) = \boldsymbol{\mathcal{L}}\, \boldsymbol{\rho}(t). \label{eq:EOM}
\end{align}
The state at time $t=0$ is $\boldsymbol{\rho}(0) \equiv \boldsymbol{\rho}_0$. Here, $\boldsymbol{\mathcal{L}}$ is the Lindblad superoperator of the two-level fluctuator (in the basis $\left\{\ketbra{+}{+},\ketbra{-}{-}\right\}$):
\begin{align}
	\boldsymbol{\mathcal{L}} = 
	\left(
	\begin{array}{cc}
	-\Gamma_+ & \Gamma_- \\
	\Gamma_+ & -\Gamma_-
	\end{array}
	\right). \label{eq:rtsLiouvillian}
\end{align}
To analyze the photon emission statistics, we resolve (or unravel) the state vector in the photon number:
\begin{align}
	\boldsymbol{\rho}(t) = \sum_{n=0}^{\infty}\boldsymbol{\ell}(n,t), \label{eq:unravelling}
\end{align}
where $\boldsymbol{\ell}(n,t)$ is the unnormalized state vector after measurement of $n$ photons in the interval $\left[0,t\right]$. More precisely, the probability of detecting $n$ photons after time $t$ is
\begin{align}
	P(n,t) = \textrm{Tr}\, \boldsymbol{\ell}(n,t),
\end{align}
where the trace $\textrm{Tr}$ is defined as the sum of all elements in the vector. Our goal is to find the measurement superoperator $\boldsymbol{\mathcal{M}}(n,t)$ so that $P(n,t)$ can be expressed as
\begin{align}
	P(n,t) = \textrm{Tr}\, \left[\boldsymbol{\mathcal{M}}(n,t) \boldsymbol{\rho}_0\right]. \label{eq:measurementProbability}
\end{align}

\begin{figure}
\centering
\includegraphics[width=\columnwidth]{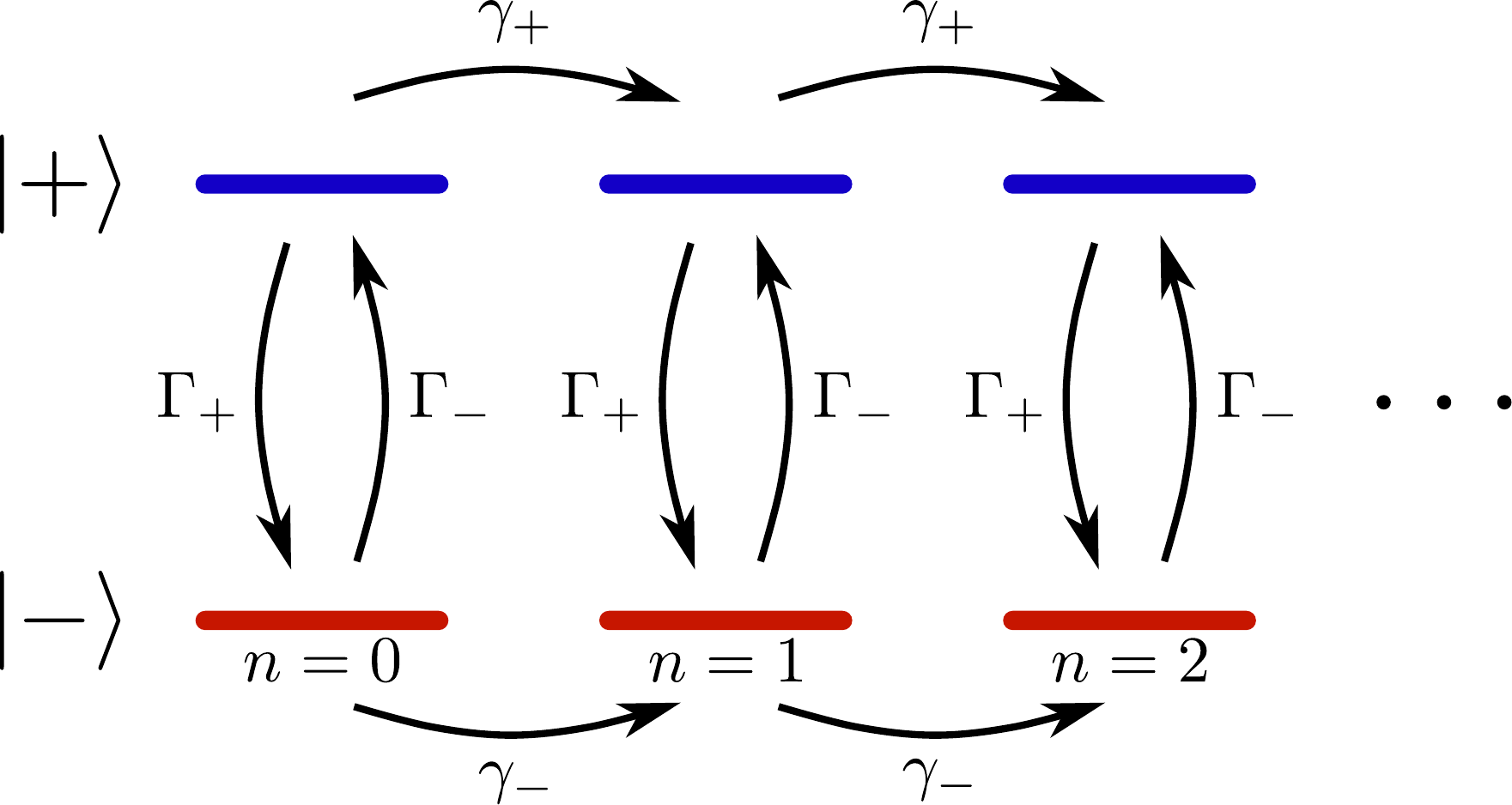}
\centering
\caption{Unraveling of the NV-center charge readout dynamics. The state $\ket{+}$ ($\ket{-}$) represents NV$^{-}$ (NV$^{0}$). Photons are detected at state-dependent rates $\gamma_{\pm}$. Each detection increases the number of observed photons $n$. The rates $\Gamma_{\pm}$ describe ionization or recombination processes that are not associated with the detection of a photon. \label{fig:fig8}}
\end{figure}
We must first obtain the equations of motion for $\boldsymbol{\ell}(n,t)$. The dynamical evolution of $\boldsymbol{\ell}(n,t)$ is illustrated in Fig.~\ref{fig:fig8}. At each time step, there are two possible types of transition. Either an ionization or recombination occurs ($n$ is unchanged) or a photon is detected ($n$ is increased by $1$). Thus, $\boldsymbol{\ell}(n,t)$ obeys the following set of coupled rate equations:
\begin{align}
	\dot{\boldsymbol{\ell}}(n,t) = \boldsymbol{\mathcal{L}}\, \boldsymbol{\ell}(n,t) - \boldsymbol{\mathcal{K}}\left[\boldsymbol{\ell}(n,t)-\boldsymbol{\ell}(n-1,t)\right]. \label{eq:unravelledEOM}
\end{align}
The initial condition is $\boldsymbol{\ell}(0,0) \equiv \boldsymbol{\ell}_0 = \boldsymbol{\rho}_0$. Here, $\boldsymbol{\mathcal{K}}$ encodes the state-dependent photon detection:
\begin{align}
	\boldsymbol{\mathcal{K}} =
	\left(
	\begin{array}{cc}
	\gamma_+ & 0 \\
	0 & \gamma_-
	\end{array}
	\right).
\end{align}
Note that summing Eq.~\eqref{eq:unravelledEOM} over all $n$ recovers Eq.~\eqref{eq:EOM} after applying Eq.~\eqref{eq:unravelling}, as required by conservation of probability. The effect of dark counts and imperfect detection efficiency is simply to modify the rates $\gamma_{\pm}$.

Equation~\eqref{eq:unravelledEOM} can be solved by introducing the characteristic function (or Fourier transform)~\cite{emary2007}
\begin{align}
\begin{split}
	&\boldsymbol{\ell}(\chi,t) = \sum_{n} \boldsymbol{\ell}(n,t)e^{i n \chi}, \\
	&\boldsymbol{\ell}(n,t) = \frac{1}{2\pi}\int_0^{2\pi}d\chi\, \boldsymbol{\ell}(\chi,t) e^{-i n \chi}, \label{eq:characteristicFunction}
\end{split}
\end{align}
where $\chi$ is a counting field. Substituting Eq.~\eqref{eq:characteristicFunction} into Eq.~\eqref{eq:unravelledEOM}, we have
\begin{align}
	\dot{\boldsymbol{\ell}}(\chi,t) = \boldsymbol{\mathcal{R}}(\chi)\,\boldsymbol{\ell}(\chi,t), \label{eq:characteristicEOM}
\end{align}
where
\begin{align}
	\boldsymbol{\mathcal{R}}(\chi) = \boldsymbol{\mathcal{L}}-\boldsymbol{\mathcal{K}} + e^{i \chi} \boldsymbol{\mathcal{K}}.
\end{align}
Solving Eq.~\eqref{eq:characteristicEOM} and inverting the characteristic function then gives
\begin{align}
	\boldsymbol{\ell}(n,t) = \boldsymbol{\mathcal{M}}(n,t) \boldsymbol{\ell}_0,
\end{align}
where the measurement superoperator $\boldsymbol{\mathcal{M}}(n,t)$ is
\begin{align}
\begin{split}
	&\boldsymbol{\mathcal{M}}(n,t) = \frac{1}{2\pi}\int_0^{2\pi} d\chi \, e^{\boldsymbol{\mathcal{R}}(\chi) t} e^{-i n \chi}, \\
	&\boldsymbol{\mathcal{M}}(n,t) = \frac{1}{n!} \left. \frac{d^n}{dz^n} e^{\boldsymbol{\mathcal{R}}(z) t} \right|_{z=0}. \label{eq:measurementOperator}
\end{split}
\end{align}
In the last line, we expressed the result in terms of the $z$-transform variable, $z \equiv e^{i \chi}$~\cite{cook1981}. We see that the measurement superoperators are generated by the matrix function $\boldsymbol{\mathcal{G}}(z,t) = e^{\boldsymbol{\mathcal{R}}(z) t}$.

We can now calculate the probability of the trajectory $\psi_N = (\delta n_{N-1}, \,\dots\, ,\delta n_0)$ given the initial state as
\begin{align}
	P(\psi_N|\boldsymbol{\ell}_0) = \textrm{Tr}\left[\mathbf{M}(\delta n_{N-1})\,\dots\,\mathbf{M}(\delta n_0)\, \boldsymbol{\ell}_0 \right], \label{eq:dataLikelihood}
\end{align}
where $\mathbf{M}(\delta n) = \boldsymbol{\mathcal{M}}(\delta n, \delta t)$ is the desired update matrix. Equation~\eqref{eq:dataLikelihood} can be used to compare the likelihood function for two arbitrary initial states $\boldsymbol{\ell}_0$, in particular the two charge states $\boldsymbol{\ell}_0^{(+)} = (1,0)^{T}$ and $\boldsymbol{\ell}_0^{(-)} = (0,1)^{T}$. If we choose a single time bin, $\delta t = t_f$, $\mathbf{M}(n)$ generates the photon number distributions given in Ref.~\cite{shields2015} (see Appendix~\ref{app:photonCounting}). Moreover, expanding $\mathbf{M}(\delta n)$ for small $\delta t$ yields the continuous-time measurement superoperator for direct detection~\cite{gambetta2001}. Note also that Eq.~\eqref{eq:dataLikelihood} is an exact solution of the continuous-time filtering equations discussed in Ref.~\cite{ng2014}.

\section{Details of Monte Carlo simulations\label{app:monteCarlo}}

In this section, we describe the Monte Carlo simulations of the error rates in more detail. For a given initial state $\boldsymbol{\rho}_0$, we generate a trajectory iteratively using Eq.~\eqref{eq:dataLikelihood} as follows:

\begin{enumerate}
	\item Knowing the state $\boldsymbol{\rho}_k$ in bin $k$, calculate the corresponding photon probability distribution $P_k(n) = \textrm{Tr}\, \left[\mathbf{M}(\delta n)\boldsymbol{\rho}_k\right]$ for $\delta n = 0,1,2,\ldots,\delta n_{\textrm{max}}$.
	\item Sample a value $\delta n_k$ at random from the distributions $P_k(\delta n)$.
	\item Find the postmeasurement state $\boldsymbol{\rho}_{k+1} = \mathbf{M}(\delta n_k) \boldsymbol{\rho}_k / P_k(\delta n_k)$.
	\item Go back to step $1$ with initial state $\boldsymbol{\rho}_{k+1}$.
\end{enumerate}
We use this procedure to simulate $1 \times 10^6$ trajectories using $\delta n_{\textrm{max}} = 5$, $\delta t = 0.1\,\textrm{ms}$, $t_M = 25\,\textrm{ms}$, and the rates extracted from experimental data (see Appendix~\ref{app:ratesExtraction}). Each trajectory is then processed using both the nonadaptive and adaptive MLE or MAP, depending on the assumed prior probabilities $P(\pm)$. 

We first simulate the nonadaptive decision rule. We choose times ranging from $t_f = 0.1\,\textrm{ms}$ to $t_f = 25\,\textrm{ms}$. For each trajectory, we use Eq.~\eqref{eq:updateRule} to calculate $\Lambda_{t_f}$. An error occurs if $\Lambda_{t_f} < \Lambda_{\textrm{th}}$ ($\Lambda_{t_f} > \Lambda_{\textrm{th}}$) when the initial state is $\ket{+}$ ($\ket{-}$). Here, $\Lambda_{\textrm{th}}=P(-)/P(+) = e^{\lambda_{\textrm{th}}}$ is the optimal decision threshold. Averaging the errors over all trajectories for each state $\ket{\pm}$ gives the conditional error rate $\epsilon^{(\pm)}$ for each time $t_f$. The error rate is then given by $\epsilon = P(+)\epsilon_+ + P(-) \epsilon_-$.

\begin{figure}
\centering
\includegraphics[width=\columnwidth]{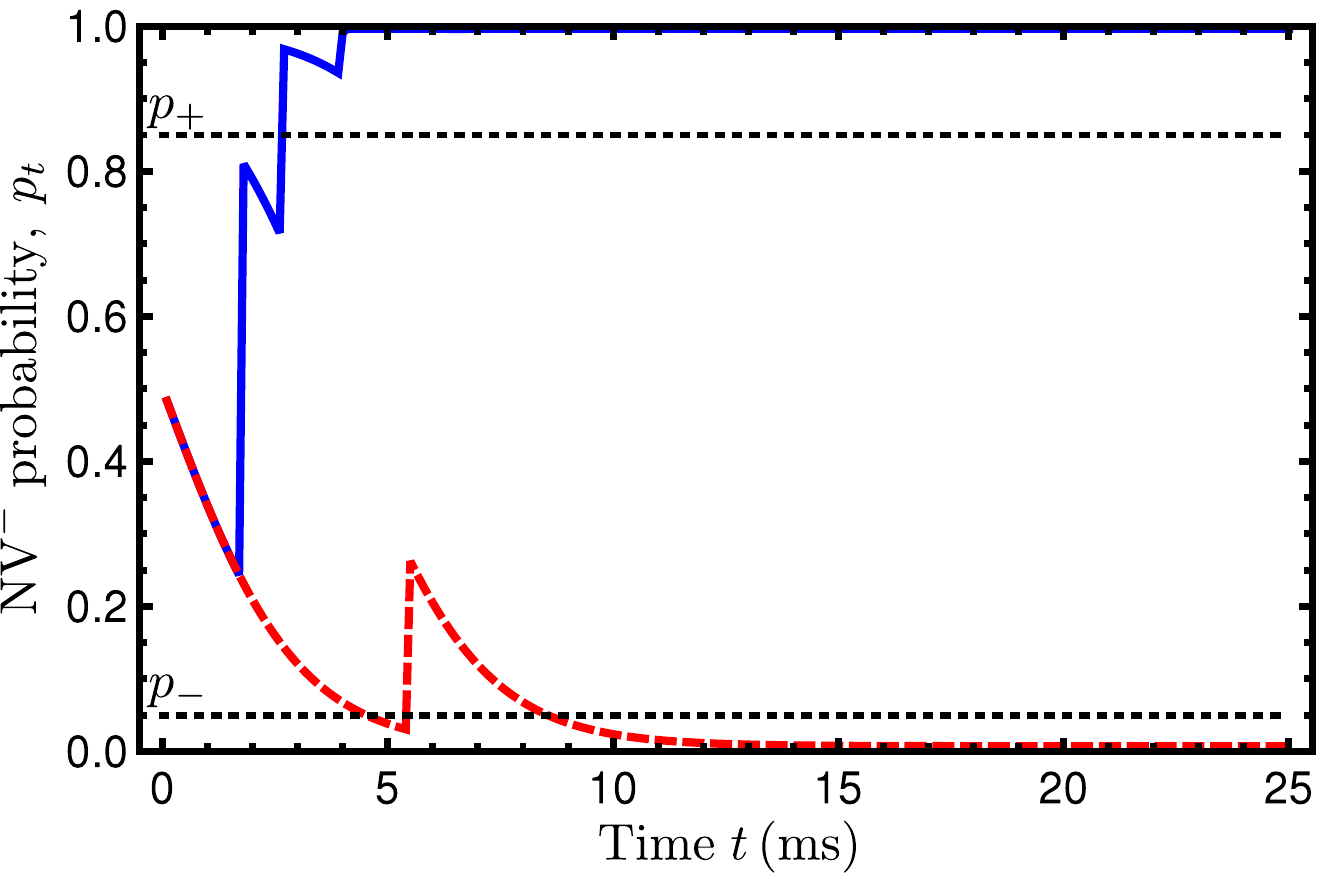}
\centering
\caption{Evolution of the NV$^{-}$ probability $p_t$ for a trajectory generated assuming $\ket{+}$ (solid blue) and $\ket{-}$ (dashed red) for the same parameters as in Fig.~\ref{fig:fig5}. In this particular case, we assumed that $P(+)=P(-)$. The discontinuities in $p_t$ correspond to the detection of a photon. Data acquisition is stopped when $p_t$ first crosses the stopping thresholds $p_+$ and $p_-$ (dotted black) or when $t$ reaches the maximum acquisition time $t_M=25\,\textrm{ms}$. \label{fig:fig9}}
\end{figure}

Next, we simulate the adaptive decision rule. We reformulate the stopping condition in terms of the stopping probabilities $p_{\pm} = e^{\lambda_{\pm} - \lambda_{\textrm{th}}}/(1+e^{\lambda_{\pm} - \lambda_{\textrm{th}}})$. Here,
\begin{align}
p_t = \frac{\Lambda_t/\Lambda_{\textrm{th}}}{1+\Lambda_t/\Lambda_{\textrm{th}}} = \frac{e^{\lambda_t - \lambda_{\textrm{th}}}}{1+e^{\lambda_t - \lambda_{\textrm{th}}}}
\end{align}
is the probability of the initial state being $\ket{+}$ when the likelihood ratio is $\Lambda_t = e^{\lambda_t}$. To vary the average readout time, we choose stopping probabilities ranging from $p_+ = 0.9$ to $p_+=1$ and from $p_- = 0$ to $p_- = 0.1$. For each trajectory, we then use Eq.~\eqref{eq:updateRule} to find the first time $t$ for which $p_t \ge p_+$, $p_t \le p_-$ or $t \ge t_M$. An error occurs if $p_t < 1/2$ ($p_t > 1/2$) when the initial state is $\ket{+}$ ($\ket{-}$). Averaging the stopping times and errors over all trajectories for each state $\ket{\pm}$ gives the conditional average readout time $T^{(\pm)}$ and the conditional error rate $\epsilon^{(\pm)}$ for each pair of stopping probabilities $\left\{p_+,p_-\right\}$. The average readout time and the error rate are calculated as $T = P(+)T^{(+)} + P(-)T^{(-)}$ and $\epsilon = P(+)\epsilon^{(+)} + P(-)\epsilon^{(-)}$, respectively. We then numerically choose the pairs of stopping probabilities that minimize the error rate under the constraint of a constant average readout time. An example of the application of the adaptive decision rule is illustrated in Fig.~\ref{fig:fig9} for both initial states.

\section{Photon-counting method \label{app:photonCounting}}

Here we summarize the analytical photon-counting readout model used in Ref.~\cite{shields2015}. In this method, the number $n$ of photons detected during the interval $\left[0,t_f\right]$ is compared to a threshold $\nu$. If $n > \nu$, we choose $\ket{+}$, while if $n \le \nu$, we choose $\ket{-}$. To calculate the error rate, we need the conditional photon distributions $P(n|\pm)$. These can be obtained within our formalism via Eq.~\eqref{eq:measurementProbability}. However, from a computational point of view, it is more efficient to use the expressions given in Ref.~\cite{shields2015}. These are
\begin{align}
	P(n|\pm) = P(n,E|\pm) + P(n,O|\pm). \label{eq:countProb}
\end{align}
Here, $E$ ($O$) corresponds to an even (odd) number of ionizations or recombinations having occurred after time $t_f$. Each term in Eq.~\eqref{eq:countProb} is given by
\begin{align}
\begin{split}
	P&(n,E|\pm) = \mathcal{P}(n,\mu_{t_f}^{(\pm)}) e^{-\Gamma_{\pm} t_f} + \\
	& \int_0^{t_f} dt\,\mathcal{P}(n,\mu_{t}^{(\pm)}) \frac{x_{t} I_1(2 x_t)}{t_f-t} e^{-\Gamma_{\pm} t} e^{-\Gamma_{\mp}(t_f-t)}, \\
	P&(n,O|\pm) = \\
	&\int_0^{t_f} dt\, \mathcal{P}(n,\mu_{t}^{(\pm)}) \Gamma_{\pm} I_0(2 x_t) e^{-\Gamma_{\pm} t} e^{-\Gamma_{\mp}(t_f-t)},
	\end{split} \label{eq:photonCountDistributions}
\end{align}
where $I_k(z)$ is the $k$th order modified Bessel function of the first kind~\cite{gradshteyn2007} and where we define
\begin{align}
\begin{split}
	&\mathcal{P}(n,\mu) = \frac{\mu^n}{n!}e^{-\mu}, \\
	& \mu_{t}^{(\pm)} = \gamma_{\pm} t + \gamma_{\mp}(t_f-t), \\
	& x_t = \sqrt{\Gamma_+ \Gamma_- t(t_f-t)}. 
\end{split} \label{eq:distributionDefinitions}
\end{align}
The conditional error rates are given by
\begin{align}
	\epsilon^{(+)} = \sum_{n \leq \nu} P(n|+), \;\; \epsilon^{(-)} = \sum_{n > \nu} P(n|-). \label{eq:photonCountConditionalErrors}
\end{align}
Substituting Eq.~\eqref{eq:photonCountDistributions} into Eq.~\eqref{eq:photonCountConditionalErrors}, we obtain
\begin{align}
	\epsilon^{(\pm)} = \epsilon_{E}^{(\pm)} + \epsilon_{O}^{(\pm)},
\end{align}
where
\begin{align}
	\begin{split}
	\epsilon &_{E}^{(\pm)} = q^{_{>}^{<}}(\nu,\mu_{t_f}^{(\pm)}) e^{-\Gamma_{\pm} t_f} + \\
	& \int_0^{t_f} dt\,q^{_{>}^{<}}(\nu,\mu_{t}^{(\pm)}) \frac{x_{t} I_1(2 x_t)}{t_f-t} e^{-\Gamma_{\pm} t} e^{-\Gamma_{\mp}(t_f-t)}, \\
	\epsilon &_{O}^{(\pm)} = \\
	&\int_0^{t_f} dt\, q^{_{>}^{<}}(\nu,\mu_{t}^{(\pm)}) \Gamma_{\pm} I_0(2 x_t) e^{-\Gamma_{\pm} t} e^{-\Gamma_{\mp}(t_f-t)}. 
	\end{split} \label{eq:photonCountErrorsExpression}
\end{align}
Here, the $q^{_{>}^{<}}$'s are cumulative distribution functions of the Poisson distribution $\mathcal{P}(n,\mu)$:
\begin{align}
\begin{split}
	&q^{<}(\nu,\mu) = Q(\nu+1,\mu), \\
	&q^{>}(\nu,\mu) = 1 - Q(\nu+1,\mu),
\end{split}
\end{align}
where $Q(a,z)$ is the regularized incomplete Gamma function~\cite{gradshteyn2007}. For a given time $t_f$, we compute the integrals in Eq.~\eqref{eq:photonCountErrorsExpression} numerically to calculate the error rate $\epsilon = P(+)\epsilon^{(+)}+P(-)\epsilon^{(-)}$ when $P(+)$ is known, or $\epsilon = \left[\epsilon^{(+)}+\epsilon^{(-)}\right]/2$ when $P(+)$ is unknown. We then increase the threshold starting from $\nu = 0$ until we find a local minimum in the error rate. We repeat the procedure at various times to plot the error rate as a function of time, as shown in Fig.~\ref{fig:fig5}.

\section{System rates and model verification \label{app:experimentalData}}

In this section, we describe how the rates $\gamma_{\pm}$ and $\Gamma_{\pm}$ were extracted from experimental data and how our readout model was validated. We first acquire $125$ photon-count trajectories, each of duration $30\,\textrm{s}$ and with time bin $\delta t = 0.1\,\textrm{ms}$. We separate the trajectories into a calibration set (60 trajectories) and a testing set (65 trajectories). The rates are extracted using only the calibration set. The testing set is then used to independently verify the validity of the charge dynamics model extracted from the calibration set.

\subsection{Extraction of the system rates \label{app:ratesExtraction}}

\begin{figure}
\centering
\includegraphics[width=\columnwidth]{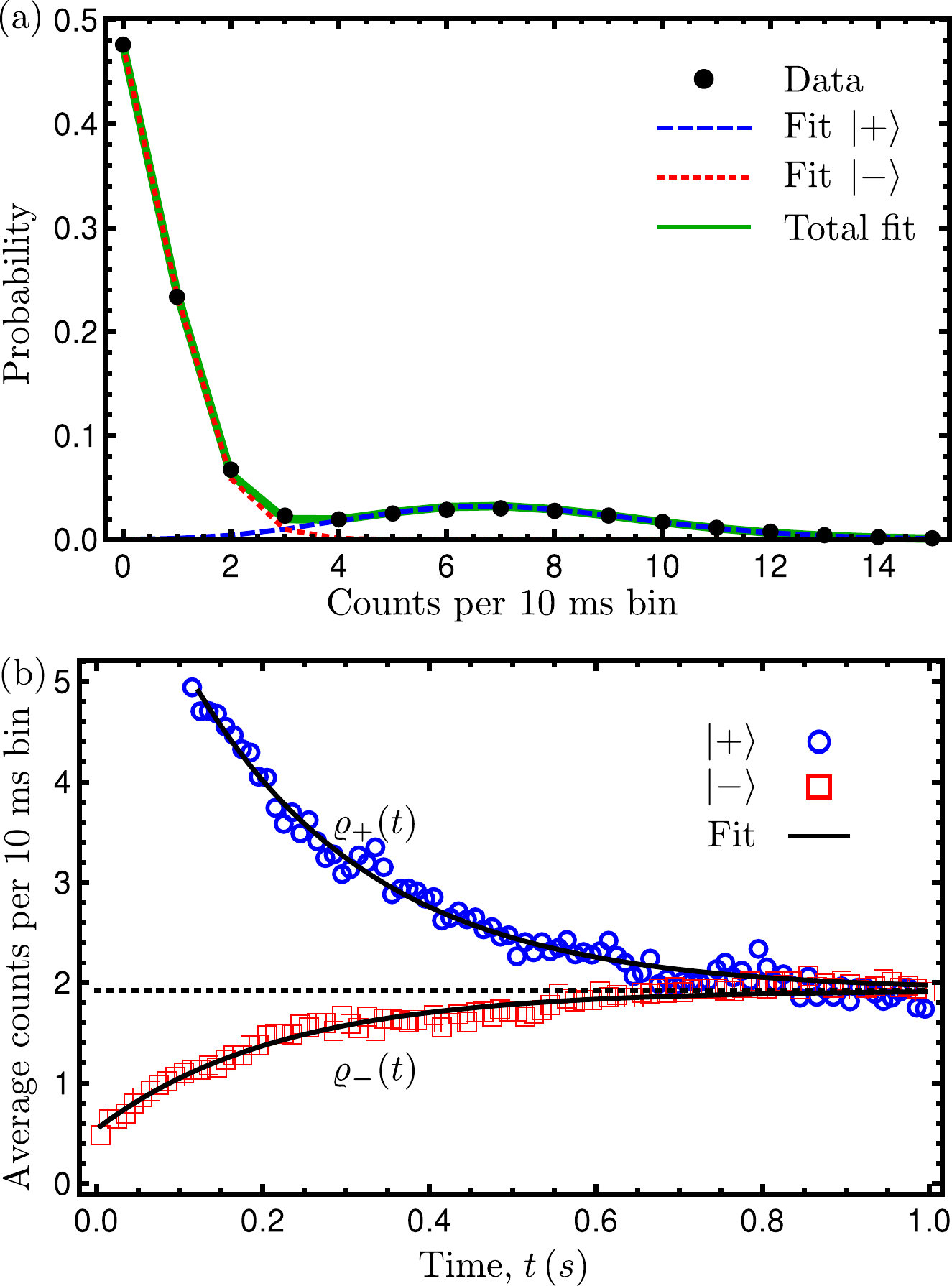}
\centering
\caption{(a) Histogram of photon counts per $10$-$\textrm{ms}$ time bin (black dots). The solid green curve is a fit to a mixture of Poisson distributions, Eq.~\eqref{eq:poissonMixture}. The dashed blue and dotted red curves are the Poisson distributions conditioned on states $\ket{+}$ and $\ket{-}$, respectively. We extract $\gamma_+ \approx 720\,\textrm{Hz}$ and $\gamma_- \approx 50\,\textrm{Hz}$ from the means of the conditional distributions. (b) Average photon counts per $10\,\textrm{ms}$, $\varrho_{\pm}(t)$, for trajectories postselected on the initial states $\ket{+}$ (blue circles) and $\ket{-}$ (red squares). The solid black lines are a simultaneous fit of $\varrho_{+}(t)$ and $\varrho_-(t)$ to Eq.~\eqref{eq:twoLevelFit}. From the fit, we extract $\Gamma_+ \approx 3.6\, \textrm{Hz}$ and $\Gamma_- \approx 0.98\, \textrm{Hz}$. The horizontal dotted black line is the steady-state average $B$ obtained from the fit to Eq.~\eqref{eq:twoLevelFit}. \label{fig:fig10}}
\end{figure}

To determine the rate $\gamma_{\pm}$, we split all trajectories of the calibration set into a total of $1800$ subtrajectories of $1\,\textrm{s}$, each of which is rebinned in bins of $\delta t = 10\,\textrm{ms}$. We then make a histogram of photon counts in all bins of $10\,\textrm{ms}$, as illustrated in Fig.~\ref{fig:fig10}(a). We fit the histogram to a mixture of Poisson distributions
\begin{align}
	P(\delta n) = \mathcal{P}(\delta n, \gamma_+ \delta t) P(+) + \mathcal{P}(\delta n, \gamma_- \delta t) P(-), \label{eq:poissonMixture}
\end{align}
where $P(\pm)$ are prior probabilities for each charge state and where $\mathcal{P}(n,\mu)$ is given in Eq.~\eqref{eq:distributionDefinitions}. The histogram is well fitted by Eq.~\eqref{eq:poissonMixture}, indicating that $\Gamma_{\pm} \delta t \ll 1$, i.e. $\Gamma_{\pm} \ll 100\,\textrm{Hz}$. From the means of the Poisson distributions, we extract $\gamma_+ \approx 720\,\textrm{Hz}$ and $\gamma_- \approx 50\,\textrm{Hz}$. When assuming no prior information on the state, the count threshold $\nu$ that best discriminates between NV$^{-}$ and NV$^{0}$ in one time bin is given by $\mathcal{P}(\nu, \gamma_+ \delta t) = \mathcal{P}(\nu, \gamma_- \delta t)$, or
\begin{align}
	\nu = \frac{\gamma_+ - \gamma_-}{\ln \left(\frac{\gamma_+}{\gamma_-}\right)} \delta t. \label{eq:poissonThreshold}
\end{align}
We find $\nu \approx 2.5$. Thus, we choose NV$^{-}$ for $\delta n > 2$ and NV$^{0}$ for $\delta n \leq 2$, with a $\approx 2 \%$ rate of error for the simplified model of Eq.~\eqref{eq:poissonMixture} (Fig.~\ref{fig:fig5} suggests that the actual error rate is closer to $3 \%$ at a measurement time of $10\,\textrm{ms}$).

To determine $\Gamma_{\pm}$, we use this thresholding procedure on the first $10$-$\textrm{ms}$ bin of each $1$-$\textrm{s}$ subtrajectory to postselect the initial state of the subtrajectory. This gives $872$ subtrajectories with the initial state identified as $\ket{+}$ and $2878$ trajectories with the initial state identified as $\ket{-}$. We then obtain the average $\varrho_{\pm}(t)$ of all trajectories for a given initial state $\ket{\pm}$, as illustrated in Fig.~\ref{fig:fig10}(b). We fit both curves simultaneously~\cite{wolk2015} to the rate equation model of the two-level fluctuator, Eq.~\eqref{eq:rtsLiouvillian}, which predicts
\begin{align}
	\varrho_{\pm}(t) = A_{\pm} e^{-\Gamma t} + B, \label{eq:twoLevelFit}
\end{align}
where
\begin{align}
	\Gamma = \Gamma_+ + \Gamma_-, \;\;\; B = \frac{\Gamma_+ (\gamma_- \delta t) + \Gamma_- (\gamma_+ \delta t)}{\Gamma_+ + \Gamma_-}.
\end{align}
From the fitted values of $\Gamma$ and $B$, we extract $\Gamma_+ \approx 3.6\, \textrm{Hz}$ and $\Gamma_- \approx 0.98\, \textrm{Hz}$. Note that in the future, the system rates could be extracted more accurately using machine learning~\cite{magesan2015}. The rates would then be chosen by training the MLE or MAP readout to minimize the empirical error rate.

\subsection{Verification of the readout model \label{app:modelVerification}}

Finally, we verify the validity of the theoretical readout model in two complementary ways using the remaining testing set and the rates extracted from the calibration set. We first compare the experimental log-likelihood ratio distribution to the theoretical prediction. We then directly calculate the error rate by preparing and subsequently measuring the charge state in postprocessing.

\begin{figure}
\centering
\includegraphics[width=\columnwidth]{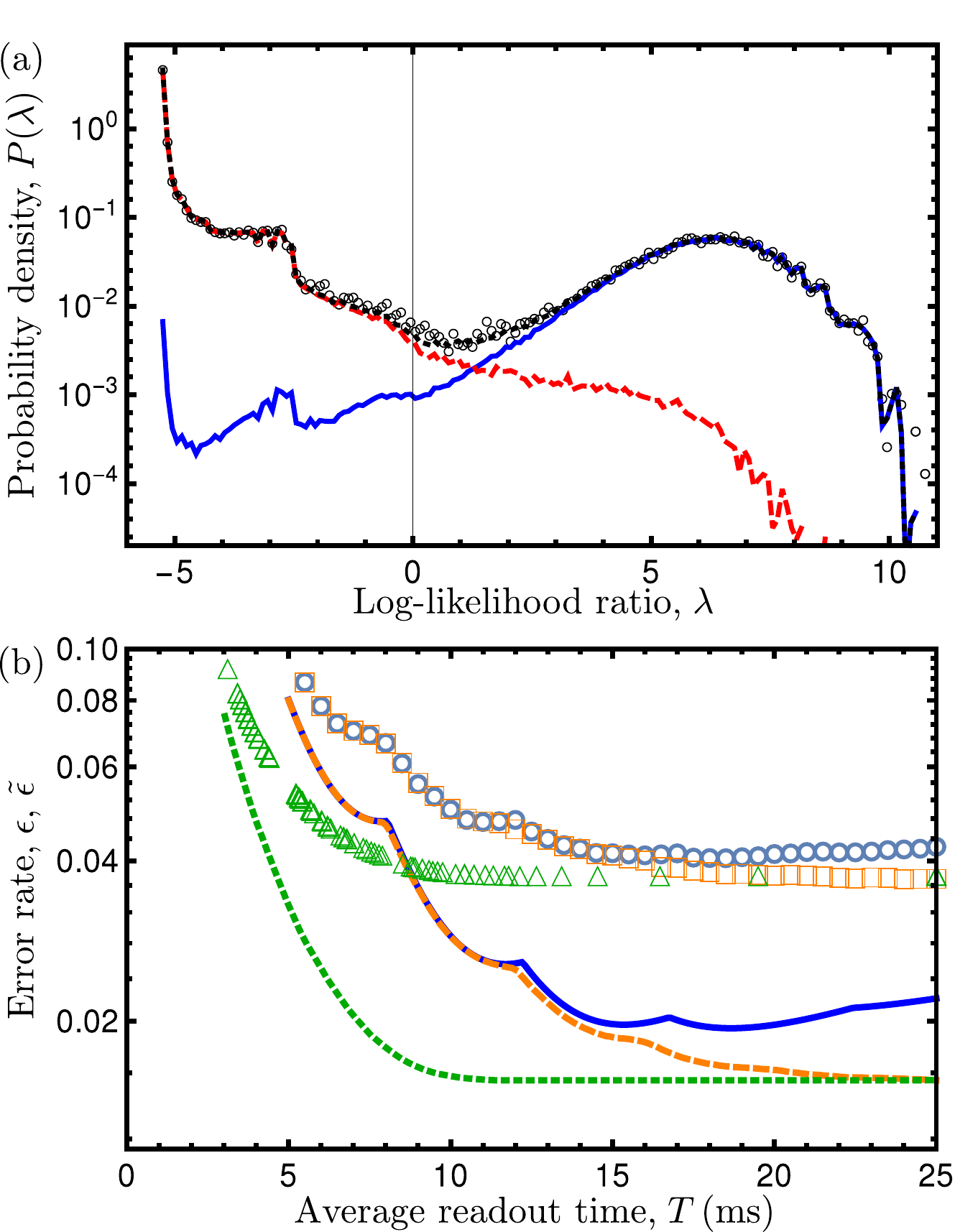}
\centering
\caption{(a) Experimentally measured probability density $P(\lambda)$ of the log-likelihood ratio $\lambda$ (black circles) for subtrajectories of duration $25\,\textrm{ms}$. The theoretical prediction $P_{\textrm{theo}}(\lambda)$ (dotted black) and the corresponding weighted conditional distributions $P_{\textrm{theo}}(\lambda|+)P_{\textrm{theo}}(+)$ (solid blue) and $P_{\textrm{theo}}(\lambda|-)P_{\textrm{theo}}(-)$ (dashed red) are shown for comparison. The theoretical prior probabilities $P_{\textrm{theo}}(\pm)$ are taken to be the stationary values corresponding to the extracted rates, Eq.~\eqref{eq:stationaryPriors}. (b) Experimental error rate $\tilde{\epsilon}$ of the photon-counting method (blue circles), the nonadaptive MLE (orange squares), and the adaptive MLE (green triangles). The theoretical error rate $\epsilon$ of Fig.~\ref{fig:fig5} is reproduced for comparison (solid blue for the photon-counting method, dashed orange for the nonadaptive MLE, and dotted green for the adaptive MLE). The discrepancy between experiment and theory can be attributed to preparation errors [see Eq.~\eqref{eq:preparationCorrection}]. \label{fig:fig11}}
\end{figure}

To calculate the experimental probability density of the log-likelihood ratio, $P(\lambda)$, we split the testing set into $78000$ subtrajectories of duration $25\,\textrm{ms}$. Using Eq.~\eqref{eq:updateRule}, we determine the value of the log-likelihood ratio $\lambda$ at the end of each subtrajectory. We use these values to calculate a histogram of $P(\lambda)$ using a bin size of $\delta \lambda$ = 0.1. The resulting distribution is shown in Fig.~\ref{fig:fig11}(a). We must compare the experimental distribution to the theoretical prediction. We first calculate the expected conditional distribution $P_{\textrm{theo}}(\lambda|\pm)$ using the Monte Carlo simulations described in Appendix~\ref{app:monteCarlo}. We then calculate the full distribution 
\begin{align}
	P_{\textrm{theo}}(\lambda) = \sum_{i=+,-} P_{\textrm{theo}}(\lambda|i)P_{\textrm{theo}}(i).
\end{align}
We assume that the prior probabilities are given by their steady-state values,
\begin{align}
P_{\textrm{theo}}(+) = 1 - P_{\textrm{theo}}(-) = \frac{\Gamma_-}{\Gamma_+ + \Gamma_-}. \label{eq:stationaryPriors}
\end{align}
The theoretical distribution is compared to the experimental distribution in Fig.~\ref{fig:fig11}(a). The prediction reproduces the experimental data without any fitting parameters. Our readout model can therefore be used to prepare the charge state with high fidelity in postselection.

To directly obtain the experimental time dependence of the error rate, we split the testing set into $39000$ subtrajectories of duration $50\,\textrm{ms}$. We prepare the state by calculating the postmeasurement state $\boldsymbol{\rho}$ after the first $25\,\textrm{ms}$ of each subtrajectory, assuming a completely mixed initial state $\boldsymbol{\rho}_0 = (1/2,1/2)^{T}$(see the procedure described at the beginning of Appendix~\ref{app:monteCarlo}). If $\rho_{+} > 1/2$ ($\rho_{+} < 1/2$), we decide that the state is most likely $\ket{+}$ ($\ket{-}$). In this way, we identify $8307$ ($30693$) subtrajectories as being in the state $\ket{+}$ ($\ket{-}$) after the first $25\,\textrm{ms}$. We then use the MLE, adaptive MLE, and photon-counting methods to read out the state using the last $25\,\textrm{ms}$ of each subtrajectory. The outcome of the readout is compared to the preparation to give the measured conditional error rates $\tilde{\epsilon}_{\pm}$. The measured error rate is then calculated as $\tilde{\epsilon} = (\tilde{\epsilon}_+ + \tilde{\epsilon}_-)/2$ and is plotted as a function of readout time in Fig.~\ref{fig:fig11}(b).

Figure~\ref{fig:fig11}(b) shows that although the adaptive-decision speedup is unchanged, the measured error rate $\tilde{\epsilon}$ is much larger than the theoretical prediction $\epsilon$. This could be the result of either imperfect preparation or imperfect modeling. We now show that this discrepancy is completely consistent with imperfect preparation. To account for preparation errors, we introduce an average preparation error rate $\eta$, assuming equal prior probabilities. The true error rate is related to the measured error rate by the relation $\tilde{\epsilon} = \eta (1-\epsilon) + (1-\eta) \epsilon$, or
\begin{align}
	\epsilon = \frac{\tilde{\epsilon}-\eta}{1-2\eta}. \label{eq:preparationCorrection}
\end{align}
We simultaneously fit the experimental error-rate curves, using $\eta$ as the only fit parameter. This gives a best-fit value of $\eta=2.22\%$. We find a good agreement between theory and experiment, as shown in Fig.~\ref{fig:fig5}. We conclude that the additional errors are likely due to preparation errors and not to imperfect modeling. We note that the transformation of Eq.~\eqref{eq:preparationCorrection} does not affect the adaptive-decision speedup.

\clearpage

\bibliographystyle{apsrev4-1-prx}
\bibliography{prx.01}

\begin{thebibliography}{53}%
\makeatletter
\providecommand \@ifxundefined [1]{%
 \@ifx{#1\undefined}
}%
\providecommand \@ifnum [1]{%
 \ifnum #1\expandafter \@firstoftwo
 \else \expandafter \@secondoftwo
 \fi
}%
\providecommand \@ifx [1]{%
 \ifx #1\expandafter \@firstoftwo
 \else \expandafter \@secondoftwo
 \fi
}%
\providecommand \natexlab [1]{#1}%
\providecommand \enquote  [1]{``#1''}%
\providecommand \bibnamefont  [1]{#1}%
\providecommand \bibfnamefont [1]{#1}%
\providecommand \citenamefont [1]{#1}%
\providecommand \href@noop [0]{\@secondoftwo}%
\providecommand \href [0]{\begingroup \@sanitize@url \@href}%
\providecommand \@href[1]{\@@startlink{#1}\@@href}%
\providecommand \@@href[1]{\endgroup#1\@@endlink}%
\providecommand \@sanitize@url [0]{\catcode `\\12\catcode `\$12\catcode
  `\&12\catcode `\#12\catcode `\^12\catcode `\_12\catcode `\%12\relax}%
\providecommand \@@startlink[1]{}%
\providecommand \@@endlink[0]{}%
\providecommand \url  [0]{\begingroup\@sanitize@url \@url }%
\providecommand \@url [1]{\endgroup\@href {#1}{\urlprefix }}%
\providecommand \urlprefix  [0]{URL }%
\providecommand \Eprint [0]{\href }%
\providecommand \doibase [0]{http://dx.doi.org/}%
\providecommand \selectlanguage [0]{\@gobble}%
\providecommand \bibinfo  [0]{\@secondoftwo}%
\providecommand \bibfield  [0]{\@secondoftwo}%
\providecommand \translation [1]{[#1]}%
\providecommand \BibitemOpen [0]{}%
\providecommand \bibitemStop [0]{}%
\providecommand \bibitemNoStop [0]{.\EOS\space}%
\providecommand \EOS [0]{\spacefactor3000\relax}%
\providecommand \BibitemShut  [1]{\csname bibitem#1\endcsname}%
\let\auto@bib@innerbib\@empty
\bibitem [{\citenamefont {Hume}\ \emph {et~al.}(2007)\citenamefont {Hume},
  \citenamefont {Rosenband},\ and\ \citenamefont {Wineland}}]{hume2007}%
  \BibitemOpen
  \bibfield  {author} {\bibinfo {author} {\bibfnamefont {D.~B.}\ \bibnamefont
  {Hume}}, \bibinfo {author} {\bibfnamefont {T.}~\bibnamefont {Rosenband}}, \
  and\ \bibinfo {author} {\bibfnamefont {D.~J.}\ \bibnamefont {Wineland}},\
  }\bibfield  {title} {\emph {\bibinfo {title} {{High-Fidelity Adaptive Qubit
  Detection through Repetitive Quantum Nondemolition Measurements}},\ }}\href
  {\doibase 10.1103/PhysRevLett.99.120502} {\bibfield  {journal} {\bibinfo
  {journal} {Phys. Rev. Lett.}\ }\textbf {\bibinfo {volume} {99}},\ \bibinfo
  {pages} {120502} (\bibinfo {year} {2007})}\BibitemShut {NoStop}%
\bibitem [{\citenamefont {Myerson}\ \emph {et~al.}(2008)\citenamefont
  {Myerson}, \citenamefont {Szwer}, \citenamefont {Webster}, \citenamefont
  {Allcock}, \citenamefont {Curtis}, \citenamefont {Imreh}, \citenamefont
  {Sherman}, \citenamefont {Stacey}, \citenamefont {Steane},\ and\
  \citenamefont {Lucas}}]{myerson2008}%
  \BibitemOpen
  \bibfield  {author} {\bibinfo {author} {\bibfnamefont {A.~H.}\ \bibnamefont
  {Myerson}}, \bibinfo {author} {\bibfnamefont {D.~J.}\ \bibnamefont {Szwer}},
  \bibinfo {author} {\bibfnamefont {S.~C.}\ \bibnamefont {Webster}}, \bibinfo
  {author} {\bibfnamefont {D.~T.~C.}\ \bibnamefont {Allcock}}, \bibinfo
  {author} {\bibfnamefont {M.~J.}\ \bibnamefont {Curtis}}, \bibinfo {author}
  {\bibfnamefont {G.}~\bibnamefont {Imreh}}, \bibinfo {author} {\bibfnamefont
  {J.~A.}\ \bibnamefont {Sherman}}, \bibinfo {author} {\bibfnamefont {D.~N.}\
  \bibnamefont {Stacey}}, \bibinfo {author} {\bibfnamefont {A.~M.}\
  \bibnamefont {Steane}}, \ and\ \bibinfo {author} {\bibfnamefont {D.~M.}\
  \bibnamefont {Lucas}},\ }\bibfield  {title} {\emph {\bibinfo {title}
  {{High-Fidelity Readout of Trapped-Ion Qubits}},\ }}\href@noop {} {\bibfield
  {journal} {\bibinfo  {journal} {Phys. Rev. Lett.}\ }\textbf {\bibinfo
  {volume} {100}},\ \bibinfo {pages} {200502} (\bibinfo {year}
  {2008})}\BibitemShut {NoStop}%
\bibitem [{\citenamefont {Burrell}\ \emph {et~al.}(2010)\citenamefont
  {Burrell}, \citenamefont {Szwer}, \citenamefont {Webster},\ and\
  \citenamefont {Lucas}}]{burrell2010}%
  \BibitemOpen
  \bibfield  {author} {\bibinfo {author} {\bibfnamefont {A.~H.}\ \bibnamefont
  {Burrell}}, \bibinfo {author} {\bibfnamefont {D.~J.}\ \bibnamefont {Szwer}},
  \bibinfo {author} {\bibfnamefont {S.~C.}\ \bibnamefont {Webster}}, \ and\
  \bibinfo {author} {\bibfnamefont {D.~M.}\ \bibnamefont {Lucas}},\ }\bibfield
  {title} {\emph {\bibinfo {title} {{Scalable Simultaneous Multiqubit Readout
  with 99.99\% Single-Shot Fidelity}},\ }}\href {\doibase
  10.1103/PhysRevA.81.040302} {\bibfield  {journal} {\bibinfo  {journal} {Phys.
  Rev. A}\ }\textbf {\bibinfo {volume} {81}},\ \bibinfo {pages} {040302}
  (\bibinfo {year} {2010})}\BibitemShut {NoStop}%
\bibitem [{\citenamefont {Noek}\ \emph {et~al.}(2013)\citenamefont {Noek},
  \citenamefont {Vrijsen}, \citenamefont {Gaultney}, \citenamefont {Mount},
  \citenamefont {Kim}, \citenamefont {Maunz},\ and\ \citenamefont
  {Kim}}]{noek2013}%
  \BibitemOpen
  \bibfield  {author} {\bibinfo {author} {\bibfnamefont {R.}~\bibnamefont
  {Noek}}, \bibinfo {author} {\bibfnamefont {G.}~\bibnamefont {Vrijsen}},
  \bibinfo {author} {\bibfnamefont {D.}~\bibnamefont {Gaultney}}, \bibinfo
  {author} {\bibfnamefont {E.}~\bibnamefont {Mount}}, \bibinfo {author}
  {\bibfnamefont {T.}~\bibnamefont {Kim}}, \bibinfo {author} {\bibfnamefont
  {P.}~\bibnamefont {Maunz}}, \ and\ \bibinfo {author} {\bibfnamefont
  {J.}~\bibnamefont {Kim}},\ }\bibfield  {title} {\emph {\bibinfo {title}
  {{High Speed, High Fidelity Detection of an Atomic Hyperfine Qubit}},\
  }}\href {\doibase 10.1364/OL.38.004735} {\bibfield  {journal} {\bibinfo
  {journal} {Opt. Lett.}\ }\textbf {\bibinfo {volume} {38}},\ \bibinfo {pages}
  {4735} (\bibinfo {year} {2013})}\BibitemShut {NoStop}%
\bibitem [{\citenamefont {Sayrin}\ \emph {et~al.}(2011)\citenamefont {Sayrin},
  \citenamefont {Dotsenko}, \citenamefont {Zhou}, \citenamefont {Peaudecerf},
  \citenamefont {Rybarczyk}, \citenamefont {Gleyzes}, \citenamefont {Rouchon},
  \citenamefont {Mirrahimi}, \citenamefont {Amini}, \citenamefont {Brune} \emph
  {et~al.}}]{sayrin2011}%
  \BibitemOpen
  \bibfield  {author} {\bibinfo {author} {\bibfnamefont {C.}~\bibnamefont
  {Sayrin}}, \bibinfo {author} {\bibfnamefont {I.}~\bibnamefont {Dotsenko}},
  \bibinfo {author} {\bibfnamefont {X.}~\bibnamefont {Zhou}}, \bibinfo {author}
  {\bibfnamefont {B.}~\bibnamefont {Peaudecerf}}, \bibinfo {author}
  {\bibfnamefont {T.}~\bibnamefont {Rybarczyk}}, \bibinfo {author}
  {\bibfnamefont {S.}~\bibnamefont {Gleyzes}}, \bibinfo {author} {\bibfnamefont
  {P.}~\bibnamefont {Rouchon}}, \bibinfo {author} {\bibfnamefont
  {M.}~\bibnamefont {Mirrahimi}}, \bibinfo {author} {\bibfnamefont
  {H.}~\bibnamefont {Amini}}, \bibinfo {author} {\bibfnamefont
  {M.}~\bibnamefont {Brune}},  \emph {et~al.},\ }\bibfield  {title} {\emph
  {\bibinfo {title} {{Real-Time Quantum Feedback Prepares and Stabilizes Photon
  Number States}},\ }}\href@noop {} {\bibfield  {journal} {\bibinfo  {journal}
  {Nature}\ }\textbf {\bibinfo {volume} {477}},\ \bibinfo {pages} {73}
  (\bibinfo {year} {2011})}\BibitemShut {NoStop}%
\bibitem [{\citenamefont {Blok}\ \emph {et~al.}(2014)\citenamefont {Blok},
  \citenamefont {Bonato}, \citenamefont {Markham}, \citenamefont {Twitchen},
  \citenamefont {Dobrovitski},\ and\ \citenamefont {Hanson}}]{blok2014}%
  \BibitemOpen
  \bibfield  {author} {\bibinfo {author} {\bibfnamefont {M.}~\bibnamefont
  {Blok}}, \bibinfo {author} {\bibfnamefont {C.}~\bibnamefont {Bonato}},
  \bibinfo {author} {\bibfnamefont {M.}~\bibnamefont {Markham}}, \bibinfo
  {author} {\bibfnamefont {D.}~\bibnamefont {Twitchen}}, \bibinfo {author}
  {\bibfnamefont {V.}~\bibnamefont {Dobrovitski}}, \ and\ \bibinfo {author}
  {\bibfnamefont {R.}~\bibnamefont {Hanson}},\ }\bibfield  {title} {\emph
  {\bibinfo {title} {{Manipulating a Qubit through the Backaction of Sequential
  Partial Measurements and Real-Time Feedback}},\ }}\href@noop {} {\bibfield
  {journal} {\bibinfo  {journal} {Nature Physics}\ }\textbf {\bibinfo {volume}
  {10}},\ \bibinfo {pages} {189} (\bibinfo {year} {2014})}\BibitemShut
  {NoStop}%
\bibitem [{\citenamefont {Murch}\ \emph {et~al.}(2015)\citenamefont {Murch},
  \citenamefont {Vijay},\ and\ \citenamefont {Siddiqi}}]{murch2015}%
  \BibitemOpen
  \bibfield  {author} {\bibinfo {author} {\bibfnamefont {K.~W.}\ \bibnamefont
  {Murch}}, \bibinfo {author} {\bibfnamefont {R.}~\bibnamefont {Vijay}}, \ and\
  \bibinfo {author} {\bibfnamefont {I.}~\bibnamefont {Siddiqi}},\ }\bibfield
  {title} {\emph {\bibinfo {title} {{Weak Measurement and Feedback in
  Superconducting Quantum Circuits}},\ }}\href@noop {} {\bibfield  {journal}
  {\bibinfo  {journal} {arXiv preprint on quant-ph/1507.04617}\ } (\bibinfo
  {year} {2015})}\BibitemShut {NoStop}%
\bibitem [{\citenamefont {Brune}\ \emph {et~al.}(2008)\citenamefont {Brune},
  \citenamefont {Bernu}, \citenamefont {Guerlin}, \citenamefont {Del\'eglise},
  \citenamefont {Sayrin}, \citenamefont {Gleyzes}, \citenamefont {Kuhr},
  \citenamefont {Dotsenko}, \citenamefont {Raimond},\ and\ \citenamefont
  {Haroche}}]{brune2008}%
  \BibitemOpen
  \bibfield  {author} {\bibinfo {author} {\bibfnamefont {M.}~\bibnamefont
  {Brune}}, \bibinfo {author} {\bibfnamefont {J.}~\bibnamefont {Bernu}},
  \bibinfo {author} {\bibfnamefont {C.}~\bibnamefont {Guerlin}}, \bibinfo
  {author} {\bibfnamefont {S.}~\bibnamefont {Del\'eglise}}, \bibinfo {author}
  {\bibfnamefont {C.}~\bibnamefont {Sayrin}}, \bibinfo {author} {\bibfnamefont
  {S.}~\bibnamefont {Gleyzes}}, \bibinfo {author} {\bibfnamefont
  {S.}~\bibnamefont {Kuhr}}, \bibinfo {author} {\bibfnamefont {I.}~\bibnamefont
  {Dotsenko}}, \bibinfo {author} {\bibfnamefont {J.~M.}\ \bibnamefont
  {Raimond}}, \ and\ \bibinfo {author} {\bibfnamefont {S.}~\bibnamefont
  {Haroche}},\ }\bibfield  {title} {\emph {\bibinfo {title} {{Process
  Tomography of Field Damping and Measurement of Fock State Lifetimes by
  Quantum Nondemolition Photon Counting in a Cavity}},\ }}\href {\doibase
  10.1103/PhysRevLett.101.240402} {\bibfield  {journal} {\bibinfo  {journal}
  {Phys. Rev. Lett.}\ }\textbf {\bibinfo {volume} {101}},\ \bibinfo {pages}
  {240402} (\bibinfo {year} {2008})}\BibitemShut {NoStop}%
\bibitem [{\citenamefont {Del{\'e}glise}\ \emph {et~al.}(2008)\citenamefont
  {Del{\'e}glise}, \citenamefont {Dotsenko}, \citenamefont {Sayrin},
  \citenamefont {Bernu}, \citenamefont {Brune}, \citenamefont {Raimond},\ and\
  \citenamefont {Haroche}}]{deleglise2008}%
  \BibitemOpen
  \bibfield  {author} {\bibinfo {author} {\bibfnamefont {S.}~\bibnamefont
  {Del{\'e}glise}}, \bibinfo {author} {\bibfnamefont {I.}~\bibnamefont
  {Dotsenko}}, \bibinfo {author} {\bibfnamefont {C.}~\bibnamefont {Sayrin}},
  \bibinfo {author} {\bibfnamefont {J.}~\bibnamefont {Bernu}}, \bibinfo
  {author} {\bibfnamefont {M.}~\bibnamefont {Brune}}, \bibinfo {author}
  {\bibfnamefont {J.-M.}\ \bibnamefont {Raimond}}, \ and\ \bibinfo {author}
  {\bibfnamefont {S.}~\bibnamefont {Haroche}},\ }\bibfield  {title} {\emph
  {\bibinfo {title} {{Reconstruction of Non-Classical Cavity Field States with
  Snapshots of their Decoherence}},\ }}\href@noop {} {\bibfield  {journal}
  {\bibinfo  {journal} {Nature}\ }\textbf {\bibinfo {volume} {455}},\ \bibinfo
  {pages} {510} (\bibinfo {year} {2008})}\BibitemShut {NoStop}%
\bibitem [{\citenamefont {Peaudecerf}\ \emph {et~al.}(2014)\citenamefont
  {Peaudecerf}, \citenamefont {Rybarczyk}, \citenamefont {Gerlich},
  \citenamefont {Gleyzes}, \citenamefont {Raimond}, \citenamefont {Haroche},
  \citenamefont {Dotsenko},\ and\ \citenamefont {Brune}}]{peaudecerf2014}%
  \BibitemOpen
  \bibfield  {author} {\bibinfo {author} {\bibfnamefont {B.}~\bibnamefont
  {Peaudecerf}}, \bibinfo {author} {\bibfnamefont {T.}~\bibnamefont
  {Rybarczyk}}, \bibinfo {author} {\bibfnamefont {S.}~\bibnamefont {Gerlich}},
  \bibinfo {author} {\bibfnamefont {S.}~\bibnamefont {Gleyzes}}, \bibinfo
  {author} {\bibfnamefont {J.~M.}\ \bibnamefont {Raimond}}, \bibinfo {author}
  {\bibfnamefont {S.}~\bibnamefont {Haroche}}, \bibinfo {author} {\bibfnamefont
  {I.}~\bibnamefont {Dotsenko}}, \ and\ \bibinfo {author} {\bibfnamefont
  {M.}~\bibnamefont {Brune}},\ }\bibfield  {title} {\emph {\bibinfo {title}
  {{Adaptive Quantum Nondemolition Measurement of a Photon Number}},\ }}\href
  {\doibase 10.1103/PhysRevLett.112.080401} {\bibfield  {journal} {\bibinfo
  {journal} {Phys. Rev. Lett.}\ }\textbf {\bibinfo {volume} {112}},\ \bibinfo
  {pages} {080401} (\bibinfo {year} {2014})}\BibitemShut {NoStop}%
\bibitem [{\citenamefont {Waldherr}\ \emph {et~al.}(2012)\citenamefont
  {Waldherr}, \citenamefont {Beck}, \citenamefont {Neumann}, \citenamefont
  {Said}, \citenamefont {Nitsche}, \citenamefont {Markham}, \citenamefont
  {Twitchen}, \citenamefont {Twamley}, \citenamefont {Jelezko},\ and\
  \citenamefont {Wrachtrup}}]{waldherr2012}%
  \BibitemOpen
  \bibfield  {author} {\bibinfo {author} {\bibfnamefont {G.}~\bibnamefont
  {Waldherr}}, \bibinfo {author} {\bibfnamefont {J.}~\bibnamefont {Beck}},
  \bibinfo {author} {\bibfnamefont {P.}~\bibnamefont {Neumann}}, \bibinfo
  {author} {\bibfnamefont {R.}~\bibnamefont {Said}}, \bibinfo {author}
  {\bibfnamefont {M.}~\bibnamefont {Nitsche}}, \bibinfo {author} {\bibfnamefont
  {M.}~\bibnamefont {Markham}}, \bibinfo {author} {\bibfnamefont
  {D.}~\bibnamefont {Twitchen}}, \bibinfo {author} {\bibfnamefont
  {J.}~\bibnamefont {Twamley}}, \bibinfo {author} {\bibfnamefont
  {F.}~\bibnamefont {Jelezko}}, \ and\ \bibinfo {author} {\bibfnamefont
  {J.}~\bibnamefont {Wrachtrup}},\ }\bibfield  {title} {\emph {\bibinfo {title}
  {{High-Dynamic-Range Magnetometry with a Single Nuclear Spin in Diamond}},\
  }}\href@noop {} {\bibfield  {journal} {\bibinfo  {journal} {Nature
  nanotechnology}\ }\textbf {\bibinfo {volume} {7}},\ \bibinfo {pages} {105}
  (\bibinfo {year} {2012})}\BibitemShut {NoStop}%
\bibitem [{\citenamefont {Shields}\ \emph {et~al.}(2015)\citenamefont
  {Shields}, \citenamefont {Unterreithmeier}, \citenamefont {de~Leon},
  \citenamefont {Park},\ and\ \citenamefont {Lukin}}]{shields2015}%
  \BibitemOpen
  \bibfield  {author} {\bibinfo {author} {\bibfnamefont {B.~J.}\ \bibnamefont
  {Shields}}, \bibinfo {author} {\bibfnamefont {Q.~P.}\ \bibnamefont
  {Unterreithmeier}}, \bibinfo {author} {\bibfnamefont {N.~P.}\ \bibnamefont
  {de~Leon}}, \bibinfo {author} {\bibfnamefont {H.}~\bibnamefont {Park}}, \
  and\ \bibinfo {author} {\bibfnamefont {M.~D.}\ \bibnamefont {Lukin}},\
  }\bibfield  {title} {\emph {\bibinfo {title} {{Efficient Readout of a Single
  Spin State in Diamond via Spin-to-Charge Conversion}},\ }}\href {\doibase
  10.1103/PhysRevLett.114.136402} {\bibfield  {journal} {\bibinfo  {journal}
  {Phys. Rev. Lett.}\ }\textbf {\bibinfo {volume} {114}},\ \bibinfo {pages}
  {136402} (\bibinfo {year} {2015})}\BibitemShut {NoStop}%
\bibitem [{\citenamefont {Elzerman}\ \emph {et~al.}(2004)\citenamefont
  {Elzerman}, \citenamefont {Hanson}, \citenamefont {{Van Beveren}},
  \citenamefont {Witkamp}, \citenamefont {Vandersypen},\ and\ \citenamefont
  {Kouwenhoven}}]{elzerman2004}%
  \BibitemOpen
  \bibfield  {author} {\bibinfo {author} {\bibfnamefont {J.~M.}\ \bibnamefont
  {Elzerman}}, \bibinfo {author} {\bibfnamefont {R.}~\bibnamefont {Hanson}},
  \bibinfo {author} {\bibfnamefont {L.~H.~W.}\ \bibnamefont {{Van Beveren}}},
  \bibinfo {author} {\bibfnamefont {B.}~\bibnamefont {Witkamp}}, \bibinfo
  {author} {\bibfnamefont {L.~M.~K.}\ \bibnamefont {Vandersypen}}, \ and\
  \bibinfo {author} {\bibfnamefont {L.~P.}\ \bibnamefont {Kouwenhoven}},\
  }\bibfield  {title} {\emph {\bibinfo {title} {{Single-Shot Read-Out of an
  Individual Electron Spin in a Quantum Dot}},\ }}\href@noop {} {\bibfield
  {journal} {\bibinfo  {journal} {Nature}\ }\textbf {\bibinfo {volume} {430}},\
  \bibinfo {pages} {431} (\bibinfo {year} {2004})}\BibitemShut {NoStop}%
\bibitem [{\citenamefont {Barthel}\ \emph {et~al.}(2009)\citenamefont
  {Barthel}, \citenamefont {Reilly}, \citenamefont {Marcus}, \citenamefont
  {Hanson},\ and\ \citenamefont {Gossard}}]{barthel2009}%
  \BibitemOpen
  \bibfield  {author} {\bibinfo {author} {\bibfnamefont {C.}~\bibnamefont
  {Barthel}}, \bibinfo {author} {\bibfnamefont {D.~J.}\ \bibnamefont {Reilly}},
  \bibinfo {author} {\bibfnamefont {C.~M.}\ \bibnamefont {Marcus}}, \bibinfo
  {author} {\bibfnamefont {M.~P.}\ \bibnamefont {Hanson}}, \ and\ \bibinfo
  {author} {\bibfnamefont {A.~C.}\ \bibnamefont {Gossard}},\ }\bibfield
  {title} {\emph {\bibinfo {title} {{Rapid Single-Shot Measurement of a
  Singlet-Triplet Qubit}},\ }}\href@noop {} {\bibfield  {journal} {\bibinfo
  {journal} {\prl}\ }\textbf {\bibinfo {volume} {103}},\ \bibinfo {pages}
  {160503} (\bibinfo {year} {2009})}\BibitemShut {NoStop}%
\bibitem [{\citenamefont {Jiang}\ \emph {et~al.}(2009)\citenamefont {Jiang},
  \citenamefont {Hodges}, \citenamefont {Maze}, \citenamefont {Maurer},
  \citenamefont {Taylor}, \citenamefont {Cory}, \citenamefont {Hemmer},
  \citenamefont {Walsworth}, \citenamefont {Yacoby}, \citenamefont {Zibrov}
  \emph {et~al.}}]{jiang2009}%
  \BibitemOpen
  \bibfield  {author} {\bibinfo {author} {\bibfnamefont {L.}~\bibnamefont
  {Jiang}}, \bibinfo {author} {\bibfnamefont {J.}~\bibnamefont {Hodges}},
  \bibinfo {author} {\bibfnamefont {J.}~\bibnamefont {Maze}}, \bibinfo {author}
  {\bibfnamefont {P.}~\bibnamefont {Maurer}}, \bibinfo {author} {\bibfnamefont
  {J.}~\bibnamefont {Taylor}}, \bibinfo {author} {\bibfnamefont
  {D.}~\bibnamefont {Cory}}, \bibinfo {author} {\bibfnamefont {P.}~\bibnamefont
  {Hemmer}}, \bibinfo {author} {\bibfnamefont {R.}~\bibnamefont {Walsworth}},
  \bibinfo {author} {\bibfnamefont {A.}~\bibnamefont {Yacoby}}, \bibinfo
  {author} {\bibfnamefont {A.}~\bibnamefont {Zibrov}},  \emph {et~al.},\
  }\bibfield  {title} {\emph {\bibinfo {title} {{Repetitive Readout of a Single
  Electronic Spin via Quantum Logic with Nuclear Spin Ancillae}},\ }}\href@noop
  {} {\bibfield  {journal} {\bibinfo  {journal} {Science}\ }\textbf {\bibinfo
  {volume} {326}},\ \bibinfo {pages} {267} (\bibinfo {year}
  {2009})}\BibitemShut {NoStop}%
\bibitem [{\citenamefont {Morello}\ \emph {et~al.}(2010)\citenamefont
  {Morello}, \citenamefont {Pla}, \citenamefont {Zwanenburg}, \citenamefont
  {Chan}, \citenamefont {Tan}, \citenamefont {Huebl}, \citenamefont
  {M{\"o}tt{\"o}nen}, \citenamefont {Nugroho}, \citenamefont {Yang},
  \citenamefont {van Donkelaar} \emph {et~al.}}]{morello2010}%
  \BibitemOpen
  \bibfield  {author} {\bibinfo {author} {\bibfnamefont {A.}~\bibnamefont
  {Morello}}, \bibinfo {author} {\bibfnamefont {J.~J.}\ \bibnamefont {Pla}},
  \bibinfo {author} {\bibfnamefont {F.~A.}\ \bibnamefont {Zwanenburg}},
  \bibinfo {author} {\bibfnamefont {K.~W.}\ \bibnamefont {Chan}}, \bibinfo
  {author} {\bibfnamefont {K.~Y.}\ \bibnamefont {Tan}}, \bibinfo {author}
  {\bibfnamefont {H.}~\bibnamefont {Huebl}}, \bibinfo {author} {\bibfnamefont
  {M.}~\bibnamefont {M{\"o}tt{\"o}nen}}, \bibinfo {author} {\bibfnamefont
  {C.~D.}\ \bibnamefont {Nugroho}}, \bibinfo {author} {\bibfnamefont
  {C.}~\bibnamefont {Yang}}, \bibinfo {author} {\bibfnamefont {J.~A.}\
  \bibnamefont {van Donkelaar}},  \emph {et~al.},\ }\bibfield  {title} {\emph
  {\bibinfo {title} {{Single-Shot Readout of an Electron Spin in Silicon}},\
  }}\href@noop {} {\bibfield  {journal} {\bibinfo  {journal} {Nature}\ }\textbf
  {\bibinfo {volume} {467}},\ \bibinfo {pages} {687} (\bibinfo {year}
  {2010})}\BibitemShut {NoStop}%
\bibitem [{\citenamefont {Neumann}\ \emph {et~al.}(2010)\citenamefont
  {Neumann}, \citenamefont {Beck}, \citenamefont {Steiner}, \citenamefont
  {Rempp}, \citenamefont {Fedder}, \citenamefont {Hemmer}, \citenamefont
  {Wrachtrup},\ and\ \citenamefont {Jelezko}}]{neumann2010}%
  \BibitemOpen
  \bibfield  {author} {\bibinfo {author} {\bibfnamefont {P.}~\bibnamefont
  {Neumann}}, \bibinfo {author} {\bibfnamefont {J.}~\bibnamefont {Beck}},
  \bibinfo {author} {\bibfnamefont {M.}~\bibnamefont {Steiner}}, \bibinfo
  {author} {\bibfnamefont {F.}~\bibnamefont {Rempp}}, \bibinfo {author}
  {\bibfnamefont {H.}~\bibnamefont {Fedder}}, \bibinfo {author} {\bibfnamefont
  {P.~R.}\ \bibnamefont {Hemmer}}, \bibinfo {author} {\bibfnamefont
  {J.}~\bibnamefont {Wrachtrup}}, \ and\ \bibinfo {author} {\bibfnamefont
  {F.}~\bibnamefont {Jelezko}},\ }\bibfield  {title} {\emph {\bibinfo {title}
  {{Single-Shot Readout of a Single Nuclear Spin}},\ }}\href@noop {} {\bibfield
   {journal} {\bibinfo  {journal} {Science}\ }\textbf {\bibinfo {volume}
  {329}},\ \bibinfo {pages} {542} (\bibinfo {year} {2010})}\BibitemShut
  {NoStop}%
\bibitem [{\citenamefont {Robledo}\ \emph
  {et~al.}(2011{\natexlab{a}})\citenamefont {Robledo}, \citenamefont
  {Childress}, \citenamefont {Bernien}, \citenamefont {Hensen}, \citenamefont
  {Alkemade},\ and\ \citenamefont {Hanson}}]{robledo2011}%
  \BibitemOpen
  \bibfield  {author} {\bibinfo {author} {\bibfnamefont {L.}~\bibnamefont
  {Robledo}}, \bibinfo {author} {\bibfnamefont {L.}~\bibnamefont {Childress}},
  \bibinfo {author} {\bibfnamefont {H.}~\bibnamefont {Bernien}}, \bibinfo
  {author} {\bibfnamefont {B.}~\bibnamefont {Hensen}}, \bibinfo {author}
  {\bibfnamefont {P.~F.}\ \bibnamefont {Alkemade}}, \ and\ \bibinfo {author}
  {\bibfnamefont {R.}~\bibnamefont {Hanson}},\ }\bibfield  {title} {\emph
  {\bibinfo {title} {{High-fidelity projective read-out of a solid-state spin
  quantum register}},\ }}\href@noop {} {\bibfield  {journal} {\bibinfo
  {journal} {Nature}\ }\textbf {\bibinfo {volume} {477}},\ \bibinfo {pages}
  {574} (\bibinfo {year} {2011}{\natexlab{a}})}\BibitemShut {NoStop}%
\bibitem [{\citenamefont {Pla}\ \emph {et~al.}(2013)\citenamefont {Pla},
  \citenamefont {Tan}, \citenamefont {Dehollain}, \citenamefont {Lim},
  \citenamefont {Morton}, \citenamefont {Zwanenburg}, \citenamefont {Jamieson},
  \citenamefont {Dzurak},\ and\ \citenamefont {Morello}}]{pla2013}%
  \BibitemOpen
  \bibfield  {author} {\bibinfo {author} {\bibfnamefont {J.~J.}\ \bibnamefont
  {Pla}}, \bibinfo {author} {\bibfnamefont {K.~Y.}\ \bibnamefont {Tan}},
  \bibinfo {author} {\bibfnamefont {J.~P.}\ \bibnamefont {Dehollain}}, \bibinfo
  {author} {\bibfnamefont {W.~H.}\ \bibnamefont {Lim}}, \bibinfo {author}
  {\bibfnamefont {J.~J.}\ \bibnamefont {Morton}}, \bibinfo {author}
  {\bibfnamefont {F.~A.}\ \bibnamefont {Zwanenburg}}, \bibinfo {author}
  {\bibfnamefont {D.~N.}\ \bibnamefont {Jamieson}}, \bibinfo {author}
  {\bibfnamefont {A.~S.}\ \bibnamefont {Dzurak}}, \ and\ \bibinfo {author}
  {\bibfnamefont {A.}~\bibnamefont {Morello}},\ }\bibfield  {title} {\emph
  {\bibinfo {title} {{High-Fidelity Readout and Control of a Nuclear Spin Qubit
  in Silicon}},\ }}\href@noop {} {\bibfield  {journal} {\bibinfo  {journal}
  {Nature}\ }\textbf {\bibinfo {volume} {496}},\ \bibinfo {pages} {334}
  (\bibinfo {year} {2013})}\BibitemShut {NoStop}%
\bibitem [{\citenamefont {Aslam}\ \emph {et~al.}(2013)\citenamefont {Aslam},
  \citenamefont {Waldherr}, \citenamefont {Neumann}, \citenamefont {Jelezko},\
  and\ \citenamefont {Wrachtrup}}]{aslam2013}%
  \BibitemOpen
  \bibfield  {author} {\bibinfo {author} {\bibfnamefont {N.}~\bibnamefont
  {Aslam}}, \bibinfo {author} {\bibfnamefont {G.}~\bibnamefont {Waldherr}},
  \bibinfo {author} {\bibfnamefont {P.}~\bibnamefont {Neumann}}, \bibinfo
  {author} {\bibfnamefont {F.}~\bibnamefont {Jelezko}}, \ and\ \bibinfo
  {author} {\bibfnamefont {J.}~\bibnamefont {Wrachtrup}},\ }\bibfield  {title}
  {\emph {\bibinfo {title} {{Photo-Induced Ionization Dynamics of the Nitrogen
  Vacancy Defect in Diamond Investigated by Single-Shot Charge State
  Detection}},\ }}\href@noop {} {\bibfield  {journal} {\bibinfo  {journal} {New
  Journal of Physics}\ }\textbf {\bibinfo {volume} {15}},\ \bibinfo {pages}
  {013064} (\bibinfo {year} {2013})}\BibitemShut {NoStop}%
\bibitem [{\citenamefont {Harty}\ \emph {et~al.}(2014)\citenamefont {Harty},
  \citenamefont {Allcock}, \citenamefont {Ballance}, \citenamefont {Guidoni},
  \citenamefont {Janacek}, \citenamefont {Linke}, \citenamefont {Stacey},\ and\
  \citenamefont {Lucas}}]{harty2014}%
  \BibitemOpen
  \bibfield  {author} {\bibinfo {author} {\bibfnamefont {T.~P.}\ \bibnamefont
  {Harty}}, \bibinfo {author} {\bibfnamefont {D.~T.~C.}\ \bibnamefont
  {Allcock}}, \bibinfo {author} {\bibfnamefont {C.~J.}\ \bibnamefont
  {Ballance}}, \bibinfo {author} {\bibfnamefont {L.}~\bibnamefont {Guidoni}},
  \bibinfo {author} {\bibfnamefont {H.~A.}\ \bibnamefont {Janacek}}, \bibinfo
  {author} {\bibfnamefont {N.~M.}\ \bibnamefont {Linke}}, \bibinfo {author}
  {\bibfnamefont {D.~N.}\ \bibnamefont {Stacey}}, \ and\ \bibinfo {author}
  {\bibfnamefont {D.~M.}\ \bibnamefont {Lucas}},\ }\bibfield  {title} {\emph
  {\bibinfo {title} {High-fidelity preparation, gates, memory, and readout of a
  trapped-ion quantum bit},\ }}\href {\doibase 10.1103/PhysRevLett.113.220501}
  {\bibfield  {journal} {\bibinfo  {journal} {Phys. Rev. Lett.}\ }\textbf
  {\bibinfo {volume} {113}},\ \bibinfo {pages} {220501} (\bibinfo {year}
  {2014})}\BibitemShut {NoStop}%
\bibitem [{\citenamefont {Jeffrey}\ \emph {et~al.}(2014)\citenamefont
  {Jeffrey}, \citenamefont {Sank}, \citenamefont {Mutus}, \citenamefont
  {White}, \citenamefont {Kelly}, \citenamefont {Barends}, \citenamefont
  {Chen}, \citenamefont {Chen}, \citenamefont {Chiaro}, \citenamefont
  {Dunsworth}, \citenamefont {Megrant}, \citenamefont {O'Malley}, \citenamefont
  {Neill}, \citenamefont {Roushan}, \citenamefont {Vainsencher}, \citenamefont
  {Wenner}, \citenamefont {Cleland},\ and\ \citenamefont
  {Martinis}}]{jeffrey2014}%
  \BibitemOpen
  \bibfield  {author} {\bibinfo {author} {\bibfnamefont {E.}~\bibnamefont
  {Jeffrey}}, \bibinfo {author} {\bibfnamefont {D.}~\bibnamefont {Sank}},
  \bibinfo {author} {\bibfnamefont {J.~Y.}\ \bibnamefont {Mutus}}, \bibinfo
  {author} {\bibfnamefont {T.~C.}\ \bibnamefont {White}}, \bibinfo {author}
  {\bibfnamefont {J.}~\bibnamefont {Kelly}}, \bibinfo {author} {\bibfnamefont
  {R.}~\bibnamefont {Barends}}, \bibinfo {author} {\bibfnamefont
  {Y.}~\bibnamefont {Chen}}, \bibinfo {author} {\bibfnamefont {Z.}~\bibnamefont
  {Chen}}, \bibinfo {author} {\bibfnamefont {B.}~\bibnamefont {Chiaro}},
  \bibinfo {author} {\bibfnamefont {A.}~\bibnamefont {Dunsworth}}, \bibinfo
  {author} {\bibfnamefont {A.}~\bibnamefont {Megrant}}, \bibinfo {author}
  {\bibfnamefont {P.~J.~J.}\ \bibnamefont {O'Malley}}, \bibinfo {author}
  {\bibfnamefont {C.}~\bibnamefont {Neill}}, \bibinfo {author} {\bibfnamefont
  {P.}~\bibnamefont {Roushan}}, \bibinfo {author} {\bibfnamefont
  {A.}~\bibnamefont {Vainsencher}}, \bibinfo {author} {\bibfnamefont
  {J.}~\bibnamefont {Wenner}}, \bibinfo {author} {\bibfnamefont {A.~N.}\
  \bibnamefont {Cleland}}, \ and\ \bibinfo {author} {\bibfnamefont {J.~M.}\
  \bibnamefont {Martinis}},\ }\bibfield  {title} {\emph {\bibinfo {title}
  {{Fast Accurate State Measurement with Superconducting Qubits}},\ }}\href
  {\doibase 10.1103/PhysRevLett.112.190504} {\bibfield  {journal} {\bibinfo
  {journal} {Phys. Rev. Lett.}\ }\textbf {\bibinfo {volume} {112}},\ \bibinfo
  {pages} {190504} (\bibinfo {year} {2014})}\BibitemShut {NoStop}%
\bibitem [{\citenamefont {Edwards}(1983)}]{edwards1983}%
  \BibitemOpen
  \bibfield  {author} {\bibinfo {author} {\bibfnamefont {A.~W.~F.}\
  \bibnamefont {Edwards}},\ }\bibfield  {title} {\emph {\bibinfo {title}
  {{Pascal's Problem: The'Gambler's Ruin'}},\ }}\href@noop {} {\bibfield
  {journal} {\bibinfo  {journal} {International Statistical Review}\ }\textbf
  {\bibinfo {volume} {51}},\ \bibinfo {pages} {73} (\bibinfo {year}
  {1983})}\BibitemShut {NoStop}%
\bibitem [{\citenamefont {Wald}(1947)}]{wald1947}%
  \BibitemOpen
  \bibfield  {author} {\bibinfo {author} {\bibfnamefont {A.}~\bibnamefont
  {Wald}},\ }\href@noop {} {\emph {\bibinfo {title} {{Sequential Analysis}}}}\
  (\bibinfo  {publisher} {Wiley},\ \bibinfo {address} {New York, U.S.A},\
  \bibinfo {year} {1947})\ \bibinfo {note} {{See the Introduction for an
  historical overview}}\BibitemShut {NoStop}%
\bibitem [{\citenamefont {Poor}(1994)}]{poor1994}%
  \BibitemOpen
  \bibfield  {author} {\bibinfo {author} {\bibfnamefont {H.~V.}\ \bibnamefont
  {Poor}},\ }\href@noop {} {\emph {\bibinfo {title} {{An Introduction to Signal
  Detection and Estimation}}}}\ (\bibinfo  {publisher} {Springer},\ \bibinfo
  {address} {Berlin, Germany},\ \bibinfo {year} {1994})\ Chap.\ \bibinfo
  {chapter} {III.D}\BibitemShut {NoStop}%
\bibitem [{\citenamefont {Risken}(1989)}]{risken1989}%
  \BibitemOpen
  \bibfield  {author} {\bibinfo {author} {\bibfnamefont {H.}~\bibnamefont
  {Risken}},\ }\href@noop {} {\emph {\bibinfo {title} {{The Fokker-Planck
  Equation}}}},\ \bibinfo {edition} {2nd}\ ed.\ (\bibinfo  {publisher}
  {Springer},\ \bibinfo {address} {Berlin, Germany},\ \bibinfo {year} {1989})\
  Chap.\ \bibinfo {chapter} {8.1}\BibitemShut {NoStop}%
\bibitem [{\citenamefont {Klebaner}(2005)}]{klebaner2005}%
  \BibitemOpen
  \bibfield  {author} {\bibinfo {author} {\bibfnamefont {F.~C.}\ \bibnamefont
  {Klebaner}},\ }\href@noop {} {\emph {\bibinfo {title} {{Introduction to
  Stochastic Calculus with Applications}}}},\ \bibinfo {edition} {2nd}\ ed.\
  (\bibinfo  {publisher} {Imperial College Press},\ \bibinfo {address} {London,
  United Kingdom},\ \bibinfo {year} {2005})\ Chap.~\bibinfo {chapter}
  {6}\BibitemShut {NoStop}%
\bibitem [{\citenamefont {Kay}(1998)}]{kay1998}%
  \BibitemOpen
  \bibfield  {author} {\bibinfo {author} {\bibfnamefont {S.~M.}\ \bibnamefont
  {Kay}},\ }\href@noop {} {\emph {\bibinfo {title} {{Fundamentals of
  Statistical Signal Processing}}}},\ Vol.~\bibinfo {volume} {II}\ (\bibinfo
  {publisher} {Prentice Hall},\ \bibinfo {address} {New Jersey, U.S.A.},\
  \bibinfo {year} {1998})\ Chap.~\bibinfo {chapter} {4}\BibitemShut {NoStop}%
\bibitem [{\citenamefont {Gambetta}\ \emph {et~al.}(2007)\citenamefont
  {Gambetta}, \citenamefont {Braff}, \citenamefont {Wallraff}, \citenamefont
  {Girvin},\ and\ \citenamefont {Schoelkopf}}]{gambetta2007}%
  \BibitemOpen
  \bibfield  {author} {\bibinfo {author} {\bibfnamefont {J.}~\bibnamefont
  {Gambetta}}, \bibinfo {author} {\bibfnamefont {W.~A.}\ \bibnamefont {Braff}},
  \bibinfo {author} {\bibfnamefont {A.}~\bibnamefont {Wallraff}}, \bibinfo
  {author} {\bibfnamefont {S.~M.}\ \bibnamefont {Girvin}}, \ and\ \bibinfo
  {author} {\bibfnamefont {R.~J.}\ \bibnamefont {Schoelkopf}},\ }\bibfield
  {title} {\emph {\bibinfo {title} {{Protocols for Optimal Readout of Qubits
  using a Continuous Quantum Nondemolition Measurement}},\ }}\href@noop {}
  {\bibfield  {journal} {\bibinfo  {journal} {Phys. Rev. A}\ }\textbf {\bibinfo
  {volume} {76}},\ \bibinfo {pages} {012325} (\bibinfo {year}
  {2007})}\BibitemShut {NoStop}%
\bibitem [{\citenamefont {Degen}(2008)}]{degen2008}%
  \BibitemOpen
  \bibfield  {author} {\bibinfo {author} {\bibfnamefont {C.}~\bibnamefont
  {Degen}},\ }\bibfield  {title} {\emph {\bibinfo {title} {{Scanning Magnetic
  Field Microscope with a Diamond Single-Spin Sensor}},\ }}\href@noop {}
  {\bibfield  {journal} {\bibinfo  {journal} {Applied Physics Letters}\
  }\textbf {\bibinfo {volume} {92}},\ \bibinfo {pages} {243111} (\bibinfo
  {year} {2008})}\BibitemShut {NoStop}%
\bibitem [{\citenamefont {Maze}\ \emph {et~al.}(2008)\citenamefont {Maze},
  \citenamefont {Stanwix}, \citenamefont {Hodges}, \citenamefont {Hong},
  \citenamefont {Taylor}, \citenamefont {Cappellaro}, \citenamefont {Jiang},
  \citenamefont {Dutt}, \citenamefont {Togan}, \citenamefont {Zibrov},
  \citenamefont {Yacoby}, \citenamefont {Walsworth},\ and\ \citenamefont
  {Lukin}}]{maze2008}%
  \BibitemOpen
  \bibfield  {author} {\bibinfo {author} {\bibfnamefont {J.}~\bibnamefont
  {Maze}}, \bibinfo {author} {\bibfnamefont {P.}~\bibnamefont {Stanwix}},
  \bibinfo {author} {\bibfnamefont {J.}~\bibnamefont {Hodges}}, \bibinfo
  {author} {\bibfnamefont {S.}~\bibnamefont {Hong}}, \bibinfo {author}
  {\bibfnamefont {J.}~\bibnamefont {Taylor}}, \bibinfo {author} {\bibfnamefont
  {P.}~\bibnamefont {Cappellaro}}, \bibinfo {author} {\bibfnamefont
  {L.}~\bibnamefont {Jiang}}, \bibinfo {author} {\bibfnamefont {M.~G.}\
  \bibnamefont {Dutt}}, \bibinfo {author} {\bibfnamefont {E.}~\bibnamefont
  {Togan}}, \bibinfo {author} {\bibfnamefont {A.}~\bibnamefont {Zibrov}},
  \bibinfo {author} {\bibfnamefont {A.}~\bibnamefont {Yacoby}}, \bibinfo
  {author} {\bibfnamefont {R.}~\bibnamefont {Walsworth}}, \ and\ \bibinfo
  {author} {\bibfnamefont {M.}~\bibnamefont {Lukin}},\ }\bibfield  {title}
  {\emph {\bibinfo {title} {{Nanoscale Magnetic Sensing with an Individual
  Electronic Spin in Diamond}},\ }}\href@noop {} {\bibfield  {journal}
  {\bibinfo  {journal} {Nature}\ }\textbf {\bibinfo {volume} {455}},\ \bibinfo
  {pages} {644} (\bibinfo {year} {2008})}\BibitemShut {NoStop}%
\bibitem [{\citenamefont {Taylor}\ \emph {et~al.}(2008)\citenamefont {Taylor},
  \citenamefont {Cappellaro}, \citenamefont {Childress}, \citenamefont {Jiang},
  \citenamefont {Budker}, \citenamefont {Hemmer}, \citenamefont {Yacoby},
  \citenamefont {Walsworth},\ and\ \citenamefont {Lukin}}]{taylor2008}%
  \BibitemOpen
  \bibfield  {author} {\bibinfo {author} {\bibfnamefont {J.}~\bibnamefont
  {Taylor}}, \bibinfo {author} {\bibfnamefont {P.}~\bibnamefont {Cappellaro}},
  \bibinfo {author} {\bibfnamefont {L.}~\bibnamefont {Childress}}, \bibinfo
  {author} {\bibfnamefont {L.}~\bibnamefont {Jiang}}, \bibinfo {author}
  {\bibfnamefont {D.}~\bibnamefont {Budker}}, \bibinfo {author} {\bibfnamefont
  {P.}~\bibnamefont {Hemmer}}, \bibinfo {author} {\bibfnamefont
  {A.}~\bibnamefont {Yacoby}}, \bibinfo {author} {\bibfnamefont
  {R.}~\bibnamefont {Walsworth}}, \ and\ \bibinfo {author} {\bibfnamefont
  {M.}~\bibnamefont {Lukin}},\ }\bibfield  {title} {\emph {\bibinfo {title}
  {{High-Sensitivity Diamond Magnetometer with Nanoscale Resolution}},\
  }}\href@noop {} {\bibfield  {journal} {\bibinfo  {journal} {Nature Physics}\
  }\textbf {\bibinfo {volume} {4}},\ \bibinfo {pages} {810} (\bibinfo {year}
  {2008})}\BibitemShut {NoStop}%
\bibitem [{\citenamefont {Balasubramanian}\ \emph {et~al.}(2008)\citenamefont
  {Balasubramanian}, \citenamefont {Chan}, \citenamefont {Kolesov},
  \citenamefont {Al-Hmoud}, \citenamefont {Tisler}, \citenamefont {Shin},
  \citenamefont {Kim}, \citenamefont {Wojcik}, \citenamefont {Hemmer},
  \citenamefont {Krueger} \emph {et~al.}}]{balasubramanian2008}%
  \BibitemOpen
  \bibfield  {author} {\bibinfo {author} {\bibfnamefont {G.}~\bibnamefont
  {Balasubramanian}}, \bibinfo {author} {\bibfnamefont {I.}~\bibnamefont
  {Chan}}, \bibinfo {author} {\bibfnamefont {R.}~\bibnamefont {Kolesov}},
  \bibinfo {author} {\bibfnamefont {M.}~\bibnamefont {Al-Hmoud}}, \bibinfo
  {author} {\bibfnamefont {J.}~\bibnamefont {Tisler}}, \bibinfo {author}
  {\bibfnamefont {C.}~\bibnamefont {Shin}}, \bibinfo {author} {\bibfnamefont
  {C.}~\bibnamefont {Kim}}, \bibinfo {author} {\bibfnamefont {A.}~\bibnamefont
  {Wojcik}}, \bibinfo {author} {\bibfnamefont {P.~R.}\ \bibnamefont {Hemmer}},
  \bibinfo {author} {\bibfnamefont {A.}~\bibnamefont {Krueger}},  \emph
  {et~al.},\ }\bibfield  {title} {\emph {\bibinfo {title} {{Nanoscale Imaging
  Magnetometry with Diamond Spins under Ambient Conditions}},\ }}\href@noop {}
  {\bibfield  {journal} {\bibinfo  {journal} {Nature}\ }\textbf {\bibinfo
  {volume} {455}},\ \bibinfo {pages} {648} (\bibinfo {year}
  {2008})}\BibitemShut {NoStop}%
\bibitem [{\citenamefont {Bechhofer}(1960)}]{bechhofer1960}%
  \BibitemOpen
  \bibfield  {author} {\bibinfo {author} {\bibfnamefont {R.}~\bibnamefont
  {Bechhofer}},\ }\bibfield  {title} {\emph {\bibinfo {title} {{A Note on the
  Limiting Relative Efficiency of the Wald Sequential Probability Ratio
  Test}},\ }}\href@noop {} {\bibfield  {journal} {\bibinfo  {journal} {Journal
  of the American Statistical Association}\ }\textbf {\bibinfo {volume} {55}},\
  \bibinfo {pages} {660} (\bibinfo {year} {1960})}\BibitemShut {NoStop}%
\bibitem [{\citenamefont {Tantaratana}\ and\ \citenamefont
  {Poor}(1982)}]{tantaratana1982}%
  \BibitemOpen
  \bibfield  {author} {\bibinfo {author} {\bibfnamefont {S.}~\bibnamefont
  {Tantaratana}}\ and\ \bibinfo {author} {\bibfnamefont {H.}~\bibnamefont
  {Poor}},\ }\bibfield  {title} {\emph {\bibinfo {title} {{Asymptotic
  Efficiencies of Truncated Sequential Tests}},\ }}\href {\doibase
  10.1109/TIT.1982.1056578} {\bibfield  {journal} {\bibinfo  {journal}
  {Information Theory, IEEE Transactions on}\ }\textbf {\bibinfo {volume}
  {28}},\ \bibinfo {pages} {911} (\bibinfo {year} {1982})}\BibitemShut
  {NoStop}%
\bibitem [{Note1()}]{Note1}%
  \BibitemOpen
  \bibinfo {note} {Other measures of state uncertainty, such as the entropy of
  the state probability distribution, may also be used when discriminating
  between more than two states~\cite {peaudecerf2014}.}\BibitemShut {Stop}%
\bibitem [{\citenamefont {D'Anjou}\ and\ \citenamefont
  {Coish}(2014)}]{danjou2014}%
  \BibitemOpen
  \bibfield  {author} {\bibinfo {author} {\bibfnamefont {B.}~\bibnamefont
  {D'Anjou}}\ and\ \bibinfo {author} {\bibfnamefont {W.~A.}\ \bibnamefont
  {Coish}},\ }\bibfield  {title} {\emph {\bibinfo {title} {{Optimal
  Post-Processing for a Generic Single-Shot Qubit Readout}},\ }}\href {\doibase
  10.1103/PhysRevA.89.012313} {\bibfield  {journal} {\bibinfo  {journal} {Phys.
  Rev. A}\ }\textbf {\bibinfo {volume} {89}},\ \bibinfo {pages} {012313}
  (\bibinfo {year} {2014})}\BibitemShut {NoStop}%
\bibitem [{\citenamefont {Khezri}\ \emph {et~al.}(2015)\citenamefont {Khezri},
  \citenamefont {Dressel},\ and\ \citenamefont {Korotkov}}]{khezri2015}%
  \BibitemOpen
  \bibfield  {author} {\bibinfo {author} {\bibfnamefont {M.}~\bibnamefont
  {Khezri}}, \bibinfo {author} {\bibfnamefont {J.}~\bibnamefont {Dressel}}, \
  and\ \bibinfo {author} {\bibfnamefont {A.~N.}\ \bibnamefont {Korotkov}},\
  }\bibfield  {title} {\emph {\bibinfo {title} {Qubit measurement error from
  coupling with a detuned neighbor in circuit qed},\ }}\href {\doibase
  10.1103/PhysRevA.92.052306} {\bibfield  {journal} {\bibinfo  {journal} {Phys.
  Rev. A}\ }\textbf {\bibinfo {volume} {92}},\ \bibinfo {pages} {052306}
  (\bibinfo {year} {2015})}\BibitemShut {NoStop}%
\bibitem [{\citenamefont {Gradshteyn}\ and\ \citenamefont
  {Ryzhik}(2007)}]{gradshteyn2007}%
  \BibitemOpen
  \bibfield  {author} {\bibinfo {author} {\bibfnamefont {I.}~\bibnamefont
  {Gradshteyn}}\ and\ \bibinfo {author} {\bibfnamefont {I.}~\bibnamefont
  {Ryzhik}},\ }\href@noop {} {\emph {\bibinfo {title} {{Table of Integrals,
  Series and Products}}}},\ \bibinfo {edition} {7th}\ ed.\ (\bibinfo
  {publisher} {Academic Press},\ \bibinfo {address} {New York, U.S.A.},\
  \bibinfo {year} {2007})\ Chap.~\bibinfo {chapter} {8}\BibitemShut {NoStop}%
\bibitem [{Note2()}]{Note2}%
  \BibitemOpen
  \bibinfo {note} {A similar process limits the readout fidelity of
  fluorescence-based readouts of trapped atoms and ions~\cite
  {gehr2010,noek2013,wolk2015}}\BibitemShut {NoStop}%
\bibitem [{\citenamefont {Gehr}\ \emph {et~al.}(2010)\citenamefont {Gehr},
  \citenamefont {Volz}, \citenamefont {Dubois}, \citenamefont {Steinmetz},
  \citenamefont {Colombe}, \citenamefont {Lev}, \citenamefont {Long},
  \citenamefont {Est\`eve},\ and\ \citenamefont {Reichel}}]{gehr2010}%
  \BibitemOpen
  \bibfield  {author} {\bibinfo {author} {\bibfnamefont {R.}~\bibnamefont
  {Gehr}}, \bibinfo {author} {\bibfnamefont {J.}~\bibnamefont {Volz}}, \bibinfo
  {author} {\bibfnamefont {G.}~\bibnamefont {Dubois}}, \bibinfo {author}
  {\bibfnamefont {T.}~\bibnamefont {Steinmetz}}, \bibinfo {author}
  {\bibfnamefont {Y.}~\bibnamefont {Colombe}}, \bibinfo {author} {\bibfnamefont
  {B.~L.}\ \bibnamefont {Lev}}, \bibinfo {author} {\bibfnamefont
  {R.}~\bibnamefont {Long}}, \bibinfo {author} {\bibfnamefont {J.}~\bibnamefont
  {Est\`eve}}, \ and\ \bibinfo {author} {\bibfnamefont {J.}~\bibnamefont
  {Reichel}},\ }\bibfield  {title} {\emph {\bibinfo {title} {{Cavity-Based
  Single Atom Preparation and High-Fidelity Hyperfine State Readout}},\ }}\href
  {\doibase 10.1103/PhysRevLett.104.203602} {\bibfield  {journal} {\bibinfo
  {journal} {Phys. Rev. Lett.}\ }\textbf {\bibinfo {volume} {104}},\ \bibinfo
  {pages} {203602} (\bibinfo {year} {2010})}\BibitemShut {NoStop}%
\bibitem [{\citenamefont {Ng}\ and\ \citenamefont {Tsang}(2014)}]{ng2014}%
  \BibitemOpen
  \bibfield  {author} {\bibinfo {author} {\bibfnamefont {S.}~\bibnamefont
  {Ng}}\ and\ \bibinfo {author} {\bibfnamefont {M.}~\bibnamefont {Tsang}},\
  }\bibfield  {title} {\emph {\bibinfo {title} {{Optimal Signal Processing for
  Continuous Qubit Readout}},\ }}\href {\doibase 10.1103/PhysRevA.90.022325}
  {\bibfield  {journal} {\bibinfo  {journal} {Phys. Rev. A}\ }\textbf {\bibinfo
  {volume} {90}},\ \bibinfo {pages} {022325} (\bibinfo {year}
  {2014})}\BibitemShut {NoStop}%
\bibitem [{\citenamefont {Gammelmark}\ \emph {et~al.}(2014)\citenamefont
  {Gammelmark}, \citenamefont {M{\o}lmer}, \citenamefont {Alt}, \citenamefont
  {Kampschulte},\ and\ \citenamefont {Meschede}}]{gammelmark2014}%
  \BibitemOpen
  \bibfield  {author} {\bibinfo {author} {\bibfnamefont {S.}~\bibnamefont
  {Gammelmark}}, \bibinfo {author} {\bibfnamefont {K.}~\bibnamefont
  {M{\o}lmer}}, \bibinfo {author} {\bibfnamefont {W.}~\bibnamefont {Alt}},
  \bibinfo {author} {\bibfnamefont {T.}~\bibnamefont {Kampschulte}}, \ and\
  \bibinfo {author} {\bibfnamefont {D.}~\bibnamefont {Meschede}},\ }\bibfield
  {title} {\emph {\bibinfo {title} {{Hidden Markov Model of Atomic Quantum Jump
  Dynamics in an Optically Probed Cavity}},\ }}\href@noop {} {\bibfield
  {journal} {\bibinfo  {journal} {Physical Review A}\ }\textbf {\bibinfo
  {volume} {89}},\ \bibinfo {pages} {043839} (\bibinfo {year}
  {2014})}\BibitemShut {NoStop}%
\bibitem [{\citenamefont {W{\"o}lk}\ \emph {et~al.}(2015)\citenamefont
  {W{\"o}lk}, \citenamefont {Piltz}, \citenamefont {Sriarunothai},\ and\
  \citenamefont {Wunderlich}}]{wolk2015}%
  \BibitemOpen
  \bibfield  {author} {\bibinfo {author} {\bibfnamefont {S.}~\bibnamefont
  {W{\"o}lk}}, \bibinfo {author} {\bibfnamefont {C.}~\bibnamefont {Piltz}},
  \bibinfo {author} {\bibfnamefont {T.}~\bibnamefont {Sriarunothai}}, \ and\
  \bibinfo {author} {\bibfnamefont {C.}~\bibnamefont {Wunderlich}},\ }\bibfield
   {title} {\emph {\bibinfo {title} {{State Selective Detection of Hyperfine
  Qubits}},\ }}\href@noop {} {\bibfield  {journal} {\bibinfo  {journal}
  {Journal of Physics B: Atomic, Molecular and Optical Physics}\ }\textbf
  {\bibinfo {volume} {48}},\ \bibinfo {pages} {075101} (\bibinfo {year}
  {2015})}\BibitemShut {NoStop}%
\bibitem [{\citenamefont {Manson}\ \emph {et~al.}(2006)\citenamefont {Manson},
  \citenamefont {Harrison},\ and\ \citenamefont {Sellars}}]{manson2006}%
  \BibitemOpen
  \bibfield  {author} {\bibinfo {author} {\bibfnamefont {N.~B.}\ \bibnamefont
  {Manson}}, \bibinfo {author} {\bibfnamefont {J.~P.}\ \bibnamefont
  {Harrison}}, \ and\ \bibinfo {author} {\bibfnamefont {M.~J.}\ \bibnamefont
  {Sellars}},\ }\bibfield  {title} {\emph {\bibinfo {title} {{Nitrogen-Vacancy
  Center in Diamond: Model of the Electronic Structure and Associated
  Dynamics}},\ }}\href {\doibase 10.1103/PhysRevB.74.104303} {\bibfield
  {journal} {\bibinfo  {journal} {Phys. Rev. B}\ }\textbf {\bibinfo {volume}
  {74}},\ \bibinfo {pages} {104303} (\bibinfo {year} {2006})}\BibitemShut
  {NoStop}%
\bibitem [{\citenamefont {Robledo}\ \emph
  {et~al.}(2011{\natexlab{b}})\citenamefont {Robledo}, \citenamefont {Bernien},
  \citenamefont {van~der Sar},\ and\ \citenamefont {Hanson}}]{robledo2011-2}%
  \BibitemOpen
  \bibfield  {author} {\bibinfo {author} {\bibfnamefont {L.}~\bibnamefont
  {Robledo}}, \bibinfo {author} {\bibfnamefont {H.}~\bibnamefont {Bernien}},
  \bibinfo {author} {\bibfnamefont {T.}~\bibnamefont {van~der Sar}}, \ and\
  \bibinfo {author} {\bibfnamefont {R.}~\bibnamefont {Hanson}},\ }\bibfield
  {title} {\emph {\bibinfo {title} {{Spin Dynamics in the Optical Cycle of
  Single Nitrogen-Vacancy Centres in Diamond}},\ }}\href@noop {} {\bibfield
  {journal} {\bibinfo  {journal} {New Journal of Physics}\ }\textbf {\bibinfo
  {volume} {13}},\ \bibinfo {pages} {025013} (\bibinfo {year}
  {2011}{\natexlab{b}})}\BibitemShut {NoStop}%
\bibitem [{\citenamefont {Doherty}\ \emph {et~al.}(2013)\citenamefont
  {Doherty}, \citenamefont {Manson}, \citenamefont {Delaney}, \citenamefont
  {Jelezko}, \citenamefont {Wrachtrup},\ and\ \citenamefont
  {Hollenberg}}]{doherty2013}%
  \BibitemOpen
  \bibfield  {author} {\bibinfo {author} {\bibfnamefont {M.~W.}\ \bibnamefont
  {Doherty}}, \bibinfo {author} {\bibfnamefont {N.~B.}\ \bibnamefont {Manson}},
  \bibinfo {author} {\bibfnamefont {P.}~\bibnamefont {Delaney}}, \bibinfo
  {author} {\bibfnamefont {F.}~\bibnamefont {Jelezko}}, \bibinfo {author}
  {\bibfnamefont {J.}~\bibnamefont {Wrachtrup}}, \ and\ \bibinfo {author}
  {\bibfnamefont {L.~C.}\ \bibnamefont {Hollenberg}},\ }\bibfield  {title}
  {\emph {\bibinfo {title} {{The Nitrogen-Vacancy Colour Centre in Diamond}},\
  }}\href@noop {} {\bibfield  {journal} {\bibinfo  {journal} {Physics Reports}\
  }\textbf {\bibinfo {volume} {528}},\ \bibinfo {pages} {1} (\bibinfo {year}
  {2013})}\BibitemShut {NoStop}%
\bibitem [{Note3()}]{Note3}%
  \BibitemOpen
  \bibinfo {note} {We note that backward equations can also be obtained for the
  case of Poissonian noise. In this case, each partial differential equation is
  replaced by a discrete set of coupled rate equations. See, for example,
  Ref.~\cite {feller1968}.}\BibitemShut {Stop}%
\bibitem [{\citenamefont {Cook}(1981)}]{cook1981}%
  \BibitemOpen
  \bibfield  {author} {\bibinfo {author} {\bibfnamefont {R.~J.}\ \bibnamefont
  {Cook}},\ }\bibfield  {title} {\emph {\bibinfo {title} {{Photon Number
  Statistics in Resonance Fluorescence}},\ }}\href@noop {} {\bibfield
  {journal} {\bibinfo  {journal} {Physical Review A}\ }\textbf {\bibinfo
  {volume} {23}},\ \bibinfo {pages} {1243} (\bibinfo {year}
  {1981})}\BibitemShut {NoStop}%
\bibitem [{\citenamefont {Emary}\ \emph {et~al.}(2007)\citenamefont {Emary},
  \citenamefont {Marcos}, \citenamefont {Aguado},\ and\ \citenamefont
  {Brandes}}]{emary2007}%
  \BibitemOpen
  \bibfield  {author} {\bibinfo {author} {\bibfnamefont {C.}~\bibnamefont
  {Emary}}, \bibinfo {author} {\bibfnamefont {D.}~\bibnamefont {Marcos}},
  \bibinfo {author} {\bibfnamefont {R.}~\bibnamefont {Aguado}}, \ and\ \bibinfo
  {author} {\bibfnamefont {T.}~\bibnamefont {Brandes}},\ }\bibfield  {title}
  {\emph {\bibinfo {title} {{Frequency-Dependent Counting Statistics in
  Interacting Nanoscale Conductors}},\ }}\href@noop {} {\bibfield  {journal}
  {\bibinfo  {journal} {Physical Review B}\ }\textbf {\bibinfo {volume} {76}},\
  \bibinfo {pages} {161404} (\bibinfo {year} {2007})}\BibitemShut {NoStop}%
\bibitem [{\citenamefont {Gambetta}\ and\ \citenamefont
  {Wiseman}(2001)}]{gambetta2001}%
  \BibitemOpen
  \bibfield  {author} {\bibinfo {author} {\bibfnamefont {J.}~\bibnamefont
  {Gambetta}}\ and\ \bibinfo {author} {\bibfnamefont {H.~M.}\ \bibnamefont
  {Wiseman}},\ }\bibfield  {title} {\emph {\bibinfo {title} {{State and
  Dynamical Parameter Estimation for Open Quantum Systems}},\ }}\href {\doibase
  10.1103/PhysRevA.64.042105} {\bibfield  {journal} {\bibinfo  {journal} {Phys.
  Rev. A}\ }\textbf {\bibinfo {volume} {64}},\ \bibinfo {pages} {042105}
  (\bibinfo {year} {2001})}\BibitemShut {NoStop}%
\bibitem [{\citenamefont {Magesan}\ \emph {et~al.}(2015)\citenamefont
  {Magesan}, \citenamefont {Gambetta}, \citenamefont {C\'orcoles},\ and\
  \citenamefont {Chow}}]{magesan2015}%
  \BibitemOpen
  \bibfield  {author} {\bibinfo {author} {\bibfnamefont {E.}~\bibnamefont
  {Magesan}}, \bibinfo {author} {\bibfnamefont {J.~M.}\ \bibnamefont
  {Gambetta}}, \bibinfo {author} {\bibfnamefont {A.~D.}\ \bibnamefont
  {C\'orcoles}}, \ and\ \bibinfo {author} {\bibfnamefont {J.~M.}\ \bibnamefont
  {Chow}},\ }\bibfield  {title} {\emph {\bibinfo {title} {{Machine Learning for
  Discriminating Quantum Measurement Trajectories and Improving Readout}},\
  }}\href {\doibase 10.1103/PhysRevLett.114.200501} {\bibfield  {journal}
  {\bibinfo  {journal} {Phys. Rev. Lett.}\ }\textbf {\bibinfo {volume} {114}},\
  \bibinfo {pages} {200501} (\bibinfo {year} {2015})}\BibitemShut {NoStop}%
\bibitem [{\citenamefont {Feller}(1968)}]{feller1968}%
  \BibitemOpen
  \bibfield  {author} {\bibinfo {author} {\bibfnamefont {W.}~\bibnamefont
  {Feller}},\ }\href@noop {} {\emph {\bibinfo {title} {{An Introduction to
  Probability Theory and Its Applications}}}},\ Vol.~\bibinfo {volume} {1}\
  (\bibinfo  {publisher} {J. Wiley \& sons},\ \bibinfo {address} {New York,
  U.S.A.},\ \bibinfo {year} {1968})\ Chap.\ \bibinfo {chapter}
  {XVII.8}\BibitemShut {NoStop}%
\end{thebibliography}%

\end{document}